\documentclass[%
 reprint,twocolumn,
nofootinbib,
 amsmath,amssymb,
 aps,
]{revtex4-2}

\usepackage{graphicx}
\usepackage{dcolumn}
\usepackage{bm}
\usepackage{xcolor}
\usepackage{hyperref}
\usepackage{cancel}

\begin{document}

\newtheorem{theorem}{Theorem}
\newtheorem{definition}{Definition}
\newtheorem{lemma}{Lemma}
\newtheorem{proposition}{Proposition}
\newtheorem{remark}{Remark}
\newtheorem{corollary}{Corollary}
\newtheorem{example}{Example}


\title{Ornstein-Uhlenbeck process and generalizations: particle's dynamics under comb constraints and stochastic resetting 
}


\author{Pece Trajanovski\textsuperscript{1}, Petar Jolakoski\textsuperscript{1}, Kiril Zelenkovski\textsuperscript{1}, Alexander Iomin\textsuperscript{2,3}, Ljupco Kocarev\textsuperscript{1,4}, Trifce Sandev\textsuperscript{1,5,6}}
\address{\textsuperscript{1}\textit{Research Center for Computer Science and Information Technologies, Macedonian Academy of Sciences and Arts, Bul. Krste Misirkov 2, 1000 Skopje, Macedonia}}
\address{\textsuperscript{2}\textit{Department of Physics, Technion, Haifa 32000, Israel}}
\address{\textsuperscript{3}Max Planck Institute for the Physics of Complex Systems, Dresden, Germany}
\address{\textsuperscript{4}\textit{Faculty of Computer Science and Engineering, Ss. Cyril and Methodius University, PO Box 393, 1000 Skopje, Macedonia}}
\address{\textsuperscript{5}Institute of Physics \& Astronomy, University of Potsdam, D-14776 Potsdam-Golm, Germany}
\address{\textsuperscript{6}\textit{Institute of Physics, Faculty of Natural Sciences and Mathematics, Ss.~Cyril and Methodius University, Arhimedova 3, 1000 Skopje, Macedonia}}

\date{\today}
\begin{abstract}
The Ornstein-Uhlenbeck process is interpreted as  Brownian motion in a harmonic potential. This  Gaussian Markov process has a bounded variance and admits a stationary probability distribution, in contrast to the standard Brownian motion.  It also tends to a drift towards its mean function, and such a process is called mean-reverting. Two examples of the generalized  Ornstein-Uhlenbeck process are considered. In the first one, we study the Ornstein-Uhlenbeck process on a comb model, as an example of the harmonically bounded random motion in the topologically constrained geometry. The main dynamical characteristics (as the first and the second moments) and the probability density function are studied in the framework of both the Langevin stochastic equation and the Fokker-Planck equation. The second example is devoted to the study of the effects of stochastic resetting on the Ornstein-Uhlenbeck process, including stochastic resetting in the comb geometry. Here, the non-equilibrium stationary state is the main question in task, where the two divergent forces, namely the resetting and the drift towards the mean, lead to compelling results both in the case of the Ornstein-Uhlenbeck process with resetting and its generalization on the two dimensional comb structure.
\end{abstract}

\maketitle

\section{Introduction}\label{secI}

Statistical treatment of Brownian motion independently suggested by Einstein~\cite{einstein} and Smoluchowski~\cite{smoluchowski} was that impact, which triggered extensive theoretical and mathematical studies of Brownian motion. In particular, Ornstein-Uhlenbeck (O-U) process is one of such phenomena. With the method introduced by Ornstein~\cite{Ornstein1919} for the velocity stochastic equation, among other remarkable results, Uhlenbeck and Ornstein were able to obtained an exact expression for the mean squared displacement (MSD) of a harmonically bound particle in Brownian motion as a function of the time and the initial deviation. The latter phenomenon is known as the O-U process~\cite{Uhlenbeck1930}. They also expressed the relation to the Fokker-Planck equation that summarised the results related to the universality  of Brownian motion, as the Markov nature phenomenon. 

In contemporary studies, it has been established that non-Markov anomalous transport is more general and ubiquitous topic across different fields of science. This issue also relates to generalization of the O-U approach to non-Markov random processes, and this generalization of the O-U process attracts much attention in many aspects from non-Markovian Langevin equations~\cite{CaBu1997} to the spectral properties of  the propagator of the Fokker-Planck equation~\cite{BeMe2006}. Nowadays, the strong motivation for studying the O-U process and its generalization relates to anomalous diffusion in inhomogeneous media leading to fractional transport~\cite{ElKl2009,Oxley2018,yuste,mardoukhi}, turbulence~\cite{Shao1995,chevillard2017}, and to its applications in the financial modeling~\cite{Maller2009}. It also includes a general aspect of a relation between a random matrix theory and Gaussian processes with long range correlations~\cite{FeKhSi2016}.

The main objective of the paper is a detailed consideration of the O-U process with and without resetting, which takes place in the comb geometry. Even though the standard and generalized O-U processes (with and without resetting) have been examined~\cite{MeSaTo2015,psingh,smith2022,SmMa2022,tweezer}, a detailed study of the influence of geometry effects, like a comb model and its anomalous properties is still an open question, which can shed light on a realization of the harmonically bounded random process in the topologically constrained geometry.

One dimensional Brownian motion affected by Poissonian resetting with a constant resetting rate is introduced in the seminal paper~\cite{Evans2011}. The issue of stochastic resetting, which is extensively explored in various diffusion processes, is well reviewed, see \textit{e.g.}, Ref.~\cite{Evans2020}. In particular, stochastic resetting is extensively explored in search processes~\cite{campos2015phase,bartumeus2009optimal,pal2020search}, population dynamics~\cite{visco2010switching}, Michaelis–Menten enzymatic reactions~\cite{reuveni2014role}, human behaviour of finding resources~\cite{bell1991behavioural}, various diffusion processes~\cite{pal2016diffusion,kusmierz2019subdiffusive,tucci2020controlling}, geometric Brownian motion
\cite{stojkoski2021autocorrelation,vinod2022nonergodicity,stojkoski2022income}, one dimensional lattices~\cite{christophorov2022resetting,bonomo2021first} and complex networks~\cite{riascos2020random,huang2021random}, as well as in quantum systems~
\cite{mukherjee2018quantum,rose2018spectral,perfetto2021designing,barkai_arxiv}, and so on. Experimental realizations of the first-passage under stochastic resetting has been demonstrated as well, using holographic optical tweezers~\cite{tal2020experimental} or laser traps~\cite{besga2020optimal}.

The continuous time random walk (CTRW) for the topologically constrained two dimensional case, known as a comb model is extensively studied and a plenty various results are well reviewed, see Refs.~\cite{Iomin2018,book_ws}, where various realizations of anomalous and heterogeneous diffusion processes are considered with an explanation of the influence of the geometry on the anomalous transport.

With these implications in mind, we suggest two main generalizations of these random processes. The first one is the problem of a diffusive particle governed by the O-U process in the comb-like structures. Here we are giving an insight on the anomalous transport
derived from the combination of anomalous diffusion, as a consequence of the comb geometry and the mean-reverting property of the process along the backbone. This interplay between the mean reverting property of the Markovian process along the backbone and Brownian motion along the fingers introduces an additional memory to the Markovian O-U process, transforming it to a very specific anomalous, non-Markovian transport in this topologically constrained geometry. The dynamics of the averaged values are studied in detail both numerically and analytically. The second problem in task is the introduction of resetting in this specific anomalous and topologically constrained O-U process. The main issue here is a creation of a non-equilibrium stationary state (NESS) by resetting inside the anomalous and stationary O-U process, which by itself is a very specific process.

Therefore, investigations of these generalizations of the O-U processes can lead to compelling results and conclusions that will be of great importance for further studies of anomalous diffusion
and its application in physics and finance, involving the O-U process, as well~\cite{ElKl2009,Oxley2018,chevillard2017,Shao1995,Maller2009}. For example, it can be helpful in description of financial models such as the models of interest rates, currency exchange rates, and commodity prices~\cite{Maller2009}.

The paper is organized as follows. In Sec.~\ref{secII}, we set the scene for the generalization of the standard O-U process. The main properties of the O-U theory are briefly discussed. In Sec.~\ref{srOU}, we present some original results on resetting in the O-U stationary transport. By observing the NESS and discussing its properties analytically and numerically, the corresponding Langevin equation is studied numerically, as well. In Sec.~\ref{OU-comb}, analytical and numerical analysis for the O-U process on the comb is suggested. We are presenting the results for the O-U particle undergoing anomalous diffusion, due to the comb-like structure, and the properties arising from that behaviour are studied. The influence of stochastic resetting on
the O-U process on the comb is investigated in Sec.~\ref{OU-comb-resetting}. In Sec.~\ref{OUfractal}, the main topological structure is a fractal grid, where the O-U process with resetting takes place. The summary of the obtained results is presented in Sec.~\ref{secSummary}. Additional information for the presented analysis on the solution to the Fokker-Planck equation of the standard O-U process
and basic definitions and relations of Hermite function, fractional integral and derivatives and the Mittag-Leffler
functions are presented in Appendices~\ref{appA}, \ref{appHermite} and~\ref{appB}, correspondingly.


\section{Ornstein-Uhlenbeck process}\label{secII}


In this section we set the scene for the generalisations of the O-U process. We do that by laying out the results for the probability density function (PDF) and the first two moments of the displacement for the standard O-U process, as well explaining the properties of the process. We define the standard O-U process in terms of the modified stochastic Langevin equation\footnote{It corresponds to the overdamped limit.}, see Refs.~\cite{Uhlenbeck1930,Doob1942, Vasicek1977},
\begin{align}\label{le str mod}
    \dot{x}(t)=\lambda \left[\mu-x(t)\right] + \sigma\,\xi(t),
\end{align}
where $\lambda$ is a parameter called rate of mean-reversion, and it represents the magnitude of the drift, $\xi(t)$ is a white noise of zero mean and correlation $\langle\xi(t)\xi(t')\rangle=\sigma\delta(t-t')$, and $\sigma$ is the variance. Larger values of $\lambda$ will cause the process to mean-revert more intensely. The parameter $\mu$ is the long-term mean value, the point to which the process is driven towards. Whenever the $x(t)$ is smaller than the long-term mean value $\mu$, the drift is positive and the process is pulling the particle towards the long-term mean, if the $x(t)$ is greater than $\mu$, the opposite happens and the drift is negative. The first part of the rhs of Eq.~(\ref{le str mod}), is the deterministic or the driving part of the process and it is what causes the mean-reversion. The second part of the rhs of the Langevin equation is the probabilistic part, due to the white noise. 

The O-U process can also be defined with its corresponding Fokker-Planck equation in the following way, 
\begin{align}\label{FPeq OUprocess} 
\frac{\partial}{\partial t}P(x,t)=L_{FP}P(x,t),
\end{align}
with the initial condition $P(x,t=0)=\delta(x-x_0)$ and zero boundary conditions at infinity both for the PDF $P(x,t)$ and its first space derivative, where
\begin{align}\label{FP_operator}
    L_{FP}\equiv\lambda\,\frac{\partial}{\partial x}(x-\mu)+\frac{\sigma^2}{2}\,\frac{\partial
^2}{\partial x^2}
\end{align}
is the Fokker-Planck operator. The solution to the partial differential equation for the O-U process (\ref{FPeq OUprocess}) can be obtained by the method of characteristics in Fourier space and it has the form, see Appendix~\ref{appA},
\begin{align}\label{pdf_solution_standard_O-U}
    P_0(x,t)=\frac{\exp\left(-\frac{\left[x-x_0e^{-\lambda t}-\mu \left(1-e^{-\lambda t}\right)\right]^2}
    {\frac{\sigma^2}{\lambda} e^{-2\lambda t}\left(e^{2\lambda t}-1\right)}
    \right)}{\sqrt{2\pi\frac{\sigma^2}{2\lambda}e^{-2\lambda t}\left(e^{2\lambda t}-1\right)}
    }.
\end{align}

Multiplying Eq.~(\ref{pdf_solution_standard_O-U}) by $x^2$ and integrating it with respect to $x$ from $-\infty$ to $\infty$ we get the differential equation for the MSD as follows 
\begin{align}
    \frac{\partial}{\partial t}\langle x^2(t)\rangle=-2 \lambda\,\langle x^2(t)\rangle +2\lambda \mu\,\langle x(t)\rangle+\sigma^2.
\end{align}
Here $\langle x(t)\rangle$ is obtained by multiplying the same equation by $x$ and integrating from $-\infty$ to $\infty$,
\begin{align}\label{mean FPeq OUprocess mod2} 
\frac{\partial}{\partial t}\langle x(t)\rangle=-\lambda\,\langle x(t)\rangle+\lambda\mu,
\end{align}
from where it follows
\begin{align}
    \langle x(t)\rangle=x_0\,e^{-\lambda t}+\mu\left(1-e^{-\lambda t}\right).
\end{align}
The final form for the MSD is
\begin{align}\label{msd standard ou mu}
    \langle x^2(t)\rangle&=x_0^2\, e^{-2\lambda t}+\mu^2\left(1+e^{-2\lambda t}\right)-2\mu^2\,e^{-\lambda t}\nonumber\\&+\frac{\sigma^2}{2\lambda}\left(1-e^{-2\lambda t}\right)+2\mu x_0\left(1-e^{-\lambda  t}\right)e^{-\lambda t}.
\end{align}
Therefore, the long time limit of the MSD saturates to $\langle x^2(t)\rangle\sim \mu^2+\frac{\sigma^2}{2\lambda}$ due to the confining potential. For $\lambda=0$, the MSD corresponds to normal diffusion, $\langle x^2(t)\rangle=x_0^2+\sigma^2t$. We also find the variance, which reads
\begin{align}
    \langle[x(t)-\langle x(t)\rangle]^2\rangle=\frac{\sigma^2}{2\lambda}\left(1-e^{-2\lambda t}\right).
\end{align}








\section{Ornstein-Uhlenbeck process with resetting}\label{srOU}

In this section we consider the problem of the O-U process in presence of stochastic Poissonian resetting~\cite{Evans2011}. This means that between two consecutive resetting events, the particle undergoes the O-U process driven towards the long term mean value $\mu$. The resetting of the particle is done to the initial position $x=x_0$, and the process is randomly repeated. The interplay of these two random phenomena results in a completely new renewal process with effects different from  the standard O-U process.

Before studying the analytic properties of the PDF of the O-U process with resetting in the framework of a renewal equation, we simulate the random trajectory by means of a discretized Langevin equation.

\subsection{Langevin equation approach}\label{LE_OU_reset}

To define the one dimensional O-U process with Poissonian resetting, we take into account the  Langevin equation~(\ref{le str mod}) for the O-U process and follow the concept of Poisonnian resetting. Namely, let us consider resetting with the rate $r$ to a fixed position. In our case, it is the initial position $x(0)=x_0$. If we suppose that at the time $t=\tau \Delta t$ the random particle is at the position $x(t)=x(\tau\Delta t)$, then for the next small  time interval $\Delta  t$ its dynamics is defined either by reset to the position $x_0$ with the probability $r\Delta t$, or by the O-U motion according to the Langevin Eq.~(\ref{le str mod}). Therefore, this dichotomous process can be simulated in the framework of the discretized Langevin equation,

\begin{align}
&x(\tau \Delta t)=\left\{
    \begin{array}{lll}\label{langevin_resetting}
    x(0), \,\,
    \textrm{with prob.}\,\, r\,\Delta t,\\ \\
    x[(\tau -1)\Delta t] + \lambda\, \big[\mu -x[(\tau -1 )\Delta t ]\big] + & \\
    \sigma\sqrt{\Delta t}\,\xi[(\tau -1)\Delta t], \,\,  \textrm{with prob.}\,\, (1-r\,\Delta t), & \end{array}
    \right.
\end{align}
where $x(0) = x_0$ is the initial particle's position. Here, we introduce the probability $r\Delta t$ for the diffusing particle to be reset to the initial position $x_0$, therefore the process is starting from the beginning, and respectively the probability $(1-r\Delta t)$ for the process to continue evolving according to the Langevin equation~(\ref{langevin_resetting})~\cite{Evans2011,Evans2020}. 
The properties of the discretized white noise are defined by zero mean $\langle \xi(\tau \Delta t) \rangle = 0$, and the correlation function $\langle \xi(\tau \Delta t)\xi(\tau'\Delta t) \rangle = \delta((\tau-\tau')\Delta t)$.

\begin{widetext}

 \begin{figure}
\centering
\includegraphics[width=17cm]{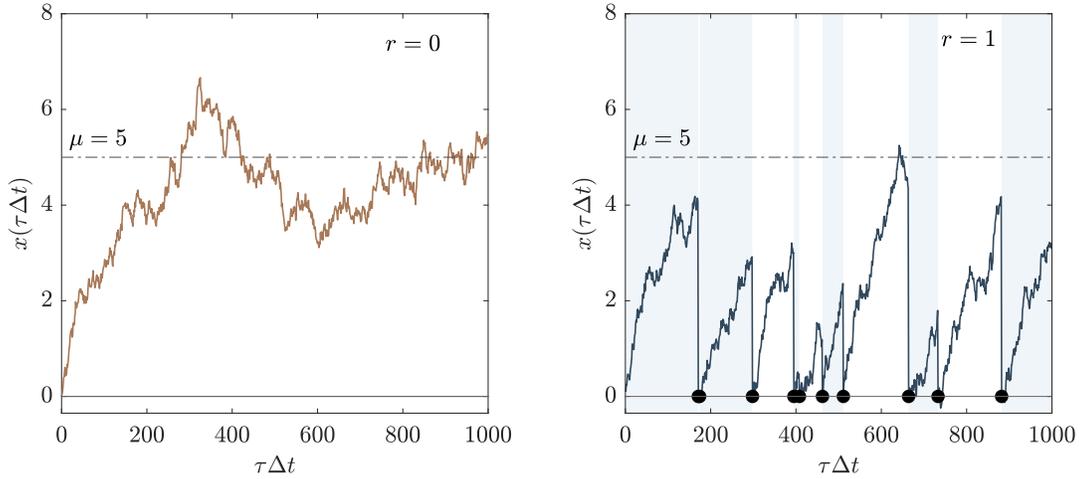}
\caption{The O-U trajectories without and with resetting ($r=1$) according to Eq.~(\ref{langevin_resetting})
for $\tau\Delta t\in (0,T)$. We set $x_0=0$, $\mu = 5$, $\lambda=1$, $\sigma=1$, $\Delta t=0.01$, $T=10^3$. 
}\label{trajectories_resett_no_resett}
\end{figure}

\end{widetext}

Results of the numerical simulations of the diffusive trajectories according to Eq.~(\ref{langevin_resetting}) without resetting (left panel with $r=0$) and with resetting (right panel with $r=1$) are presented in Fig.~\ref{trajectories_resett_no_resett}. The O-U trajectory tends to random oscillations around its  long-term mean value $\mu$, while random resets change the trajectory drastically. That eventually leads to a new equation with the solution for the PDF 
$P_r(x,t)$.


\subsection{Probability density function and non-equilibrium stationary state}

From the Langevin description of the O-U process with stochastic resetting one can find the governing Fokker-Planck equation, which reads 
\begin{align}\label{FPeq OUprocess reset} 
\frac{\partial}{\partial t}P_r(x,t)=L_{FP}P_r(x,t)-r\,P_r(x,t)+r\,\delta(x-x_0),
\end{align}
with the initial condition $P_r(x,t=0)=P(x,t=0)=\delta(x-x_0)$ and zero boundary conditions at infinity. Here, $L_{FP}$ is defined in Eq.~(\ref{FP_operator}) and $-r$ is the loss of the probability at the
position $x$ due to the reset to the initial position $x=x_0$, while the gain of the probability takes place with the rate $+r$ at the initial position $x_0$.


To find the solution to Eq.~(\ref{FPeq OUprocess reset}), the Laplace transform is applied,
$\mathcal{L}[P_r(x,t)]=\int_{0}^{\infty}e^{-st}P_r(x,t)\,dt=\hat{P}_r(x,s)$, 
which yields the expression
\begin{align}
    s\,\hat{P}_r (x,s)-\delta (x-x_0)=\frac{s}
{s+r}\,L_{FP}\hat{P}_r(x,s).
\end{align}
Then the inverse Laplace transform yields Eq.(\ref{FPeq OUprocess reset}) in the form
\begin{align}\label{FEq_OU_reset_kernel}
    \frac{\partial}{\partial t} P_r (x,t)=\frac{d}{d t}\int_{0}^{t}\eta(t-t')\,
    L_{FP}P_r (x,t')\,dt'.
\end{align}
Here $\eta(t)=e^{-r t}$ and $\hat{\eta}(s)=\frac{1}{s+r}$. This equation can be solved by using the subordination approach~\cite{Metzler2000,
Magdziarz2009,
Barkai2001,
Meerschaert2002,
Bazhlekova2019}. 
Then, we look for the solution of Eq.~(\ref{FEq_OU_reset_kernel}) in the form of the subordination integral
\begin{align}\label{subordination}
    P_r(x,t)=\int_{0}^{\infty} P_0(x,u)\,h(u,t)\,du,
\end{align}
where $P_0(x,y)$ is the O-U solution of Eq.~(\ref{FPeq OUprocess}), and $h(u,t)$ is the  subordination function. Now by the Laplace transform of Eq.(\ref{subordination}), and by using the subordination function 
\begin{align}
    \hat{h}(u,s)=\frac{1}{s\, \hat{\eta} (s)}e^{-\frac{u}{\hat{\eta}(s)}},
\end{align}
we obtain the following expression for the Laplace image of the PDF

\begin{align}\label{P_r-P_o}
    \hat{P}_r (x,s)&=\int_{0}^{\infty} P_0(x,u)\,\hat{h}(u,s)\,du\nonumber\\&=\frac{1}{s\, \hat{\eta} (s)}\int_{0}^{\infty} P_0(x,u)\,e^{-\frac{u}{\hat{\eta}(s)}}\,du\nonumber\\&=\frac{1}{s\, \hat{\eta}(s)}\, \hat{P}_0\left(x,1/\hat{\eta}(s)\right)=\frac{s+r}{s}\,\hat{P}_{0}(x,s+r).
\end{align}
The inverse Laplace transform yields the following renewal equation~\cite{evans2014,Evans2020,mendez2019,bodrova}
\begin{align}\label{pdf renewal ou}
    P_{r}(x,t)&=e^{-rt}\,P_0(x,t)+\int_{0}^{t}r\,e^{-rt'}\,P_0(x,t')\,dt'.
\end{align}

\begin{widetext}

\begin{figure}[h!]
\centering
\includegraphics[width=15cm]{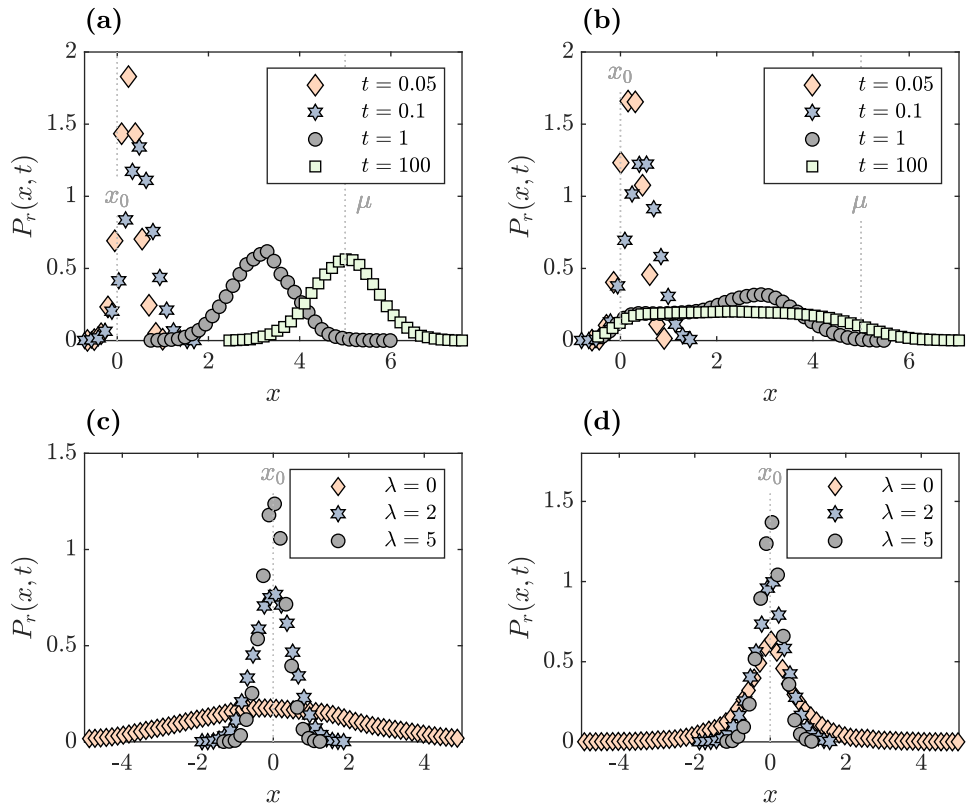}
\caption{Simulations of the PDF of the O-U process according to the Langevin equation~(\ref{langevin_resetting}); \textbf{(a)} Evolution of the PDF for $x_0=0$, $r=0$, $\mu = 5$, $\lambda=1$, $\sigma=1$, $\Delta t=0.01$, for an initial ensemble of $N=10^4$ trajectories; \textbf{(b)} Same as (a) with the resetting rate $r=1$; \textbf{(c)} PDF for different values of the rate of mean-reversion $\lambda$ and $x_0=0$, $t=5$, $\mu = 0$, $\sigma=1$, $\Delta t=0.01$, $N=10^4$ with resetting rate r=0; \textbf{(d)} Same as (c) with the resetting rate $r=1$.}\label{PDF_no_comb_plot}
\end{figure}

\end{widetext}

The temporal evolution of the PDFs  without, and with resetting according to Eq.~(\ref{pdf renewal ou}) are presented in Figs.~\ref{PDF_no_comb_plot}~(a) and~(b), respectively. The simulated PDFs for the different values of the mean-reverting rate at time $t = 5$ without and with resetting are shown in Figs.~\ref{PDF_no_comb_plot}~(c) and~(d), respectively.


From Eq.~(\ref{P_r-P_o}), we obtain that in the long time limit the system reaches a NESS, given by
\begin{align}\label{ness standard ou}
    P^{st}_{r}(x)=\lim_{t\rightarrow\infty}P_r (x,t)=\lim_{s\rightarrow0}s\,\hat{P}_r (x,s)=r\,\hat{P}_{0}(x,r).
\end{align}
Note that in the limit $t\rightarrow \infty$ the time derivative in Eq. (8) tends to zero, that is, $\frac{\partial}{\partial t}P_r(x,t)=0$, which yields
\begin{align}\label{FPeq OUprocess reset_stationary} 0=L_{FP}P_r^{st}(x)-r\,P_r^{st}(x)+r\,\delta(x-x_0).
\end{align}

The solution to Eq.~(\ref{FPeq OUprocess reset_stationary}), which is the NESS, is obtained as follows. Following the standard procedure, we consider two regions $x>x_0$ and $x<x_0$. Therefore, the corresponding solution for $x>x_0$ is $P_{r,1}^{st}(x)$ while when $x<x_0$ the solution is $P_{r,2}^{st}(x)$. The solution should be continuous at $x=x_0$, i.e.,
\begin{align}
    P_{r,1}^{st}(x)|_{x=x_0}=P_{r,2}^{st}(x)|_{x=x_0}.
\end{align} 
Moreover, by integration of Eq.~(\ref{FPeq OUprocess reset_stationary}) in vicinity of $x=x_0$, one finds
\begin{align}
    \left.\frac{d}{dx}P_{r,1}^{st}(x)\right|_{x=x_0}-\left.\frac{d}{dx}P_{r,2}^{st}(x)\right|_{x=x_0}=-\frac{r}{\sigma^2/2},
\end{align}
which means that the first derivatives at $x=x_0$ have a discontinuity. Let us first consider the equation for $x>x_0$, \begin{align}\label{FPeq OUprocess reset x>x0} 
0=\lambda\,\frac{\partial}{\partial x}\left[(x-\mu)\,P_{r,1}^{st}(x)\right]+\frac{\sigma^2}{2}\,\frac{\partial
^2}{\partial x^2}P_{r,2}^{st}(x)-r\,P_{r,1}^{st}(x).
\end{align}
Using Mathematica, we obtain the solution as follows
\begin{align}\label{sol pr1}
    P_{r,1}^{st}(x)&=c_1\, e^{-\frac{\lambda x(x-2\mu)}{\sigma^2}}\,H_{-\frac{r}{\lambda}}\left(\frac{\sqrt{\lambda}(x-\mu)}{\sigma}\right)\nonumber\\&+c_2\, e^{-\frac{\lambda x(x-2\mu)}{\sigma^2}}\,{_1}F_{1}\left(\frac{r}{2\lambda},\frac{1}{2},\frac{\lambda (x-\mu)^2}{\sigma^2}\right),
\end{align}
where $c_{1,2}$ are constants, $H_{\nu}(z)$ is the Hermite function (for details, see Appendix~\ref{appHermite}), while ${_1}F_{1}(a,b,z)$ is the confluent hypergeometric function. For $x<x_0$, we have
\begin{align}\label{FPeq OUprocess reset x<x0} 
0=\lambda\,\frac{\partial}{\partial x}\left[(x-\mu)\,P_{r,2}^{st}(x)\right]+\frac{\sigma^2}{2}\,\frac{\partial
^2}{\partial x^2}P_{r,2}^{st}(x)-r\,P_{r,2}^{st}(x),
\end{align}
and the solution reads 
\begin{align}\label{sol pr2}
    P_{r,2}^{st}(x)&=c_3\, e^{-\frac{\lambda x(x-2\mu)}{\sigma^2}}\,H_{-\frac{r}{\lambda}}\left(\frac{\sqrt{\lambda}(x-\mu)}{\sigma}\right)\nonumber\\&+c_4\, e^{-\frac{\lambda x(x-2\mu)}{\sigma^2}}\,{_1}F_{1}\left(\frac{r}{2\lambda},\frac{1}{2},\frac{\lambda (x-\mu)^2}{\sigma^2}\right),
\end{align}
where $c_{3,4}$ are constants. 
One should also take into consideration the normalization condition
\begin{align}
    \int_{-\infty}^{x_0}P_{r,2}^{st}(x)\,dx+\int_{x_0}^{-\infty}P_{r,1}^{st}(x)\,dx=1.
\end{align}
Since the obtained solutions are too complicated for the  analytical evaluation of the coefficients, a numerical procedure is suggested. Note also that in the case of $\mu=0$, we refer to the results obtained  by Pal~\cite{pal2015diffusion}.

\begin{figure}[h!]
\includegraphics[width=9cm]{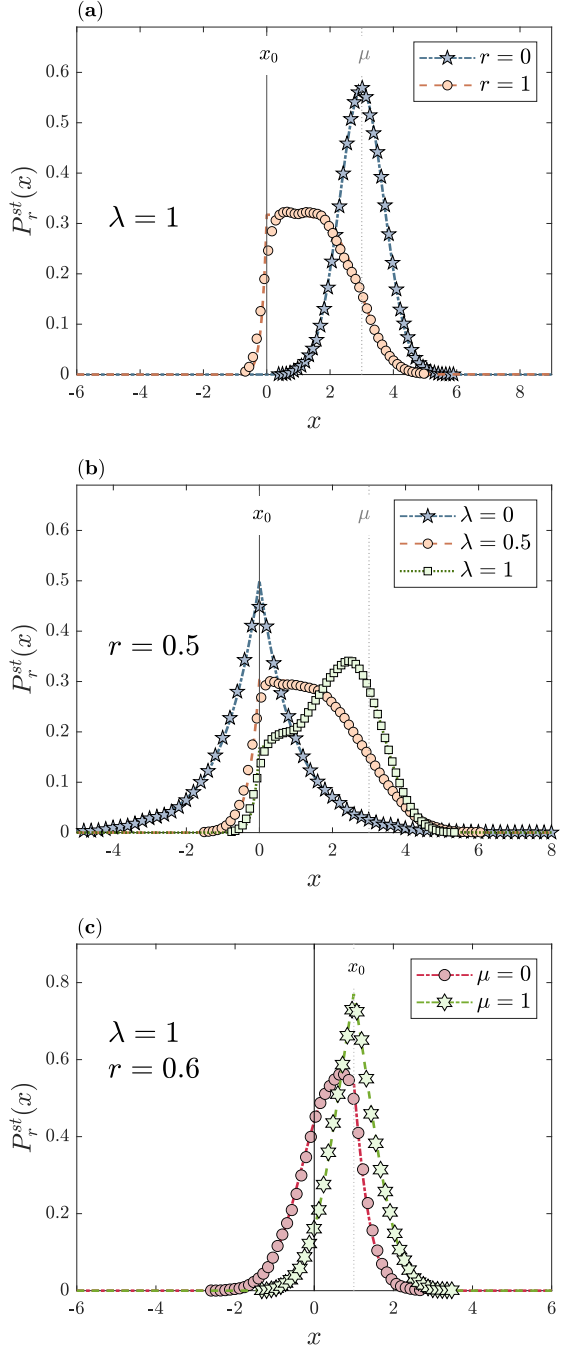}
\vspace{-0.5cm}
\caption{Numerical results (lines) and simulations (markers) of the NESS~(\ref{ness standard ou}): 
(a) for different values of $r=\{0, 1\}$; (b) different values of $\lambda=\{0, 0.5, 1\}$ with parameters: $\sigma=1$, $\mu=3$, $dt=10^{-3}$, $r=0.5$, $x_0=0$ and an initial ensemble of $N=10^4$ trajectories; (c) we set $x_0=1$, $\sigma=1$, $\mu=\{ 0, 1\}$, $\lambda=1$, $dt=10^{-3}$, $r=0.6$, and an initial ensemble of $N=10^4$ trajectories --- for $\mu=0$, we get the same results given by Pal~\cite{pal2015diffusion}.}\label{Stationary_distribution_different_lambda}
\end{figure}

Therefore, the NESS, which is the solution of Eq.~(\ref{FPeq OUprocess reset_stationary}), is computed numerically and confirmed with Monte-Carlo simulations. The methods used here for the numerical computation of the PDF are the ``shooting and 4th order Runge-Kutta methods'', see Ref.~\cite{Bailey1968}. The ``shooting method'' is used for approximating boundary-value problems by initial value problems. With this method, the missing initial conditions are guessed and then the 4th order Runge-Kutta method is used for solving the approximated initial value problem. The results of the numerical calculations are presented in Fig.~\ref{Stationary_distribution_different_lambda},
where the numerical results obtained by the ``shooting and 4th order Runge-Kutta methods'' are presented by lines, while the simulation results are presented by 
markers. The NESS, as the PDF, are obtained by the Monte-Carlo simulation of the Langevin equation~(\ref{langevin_resetting}) for different values of the resetting rate $r$, see Fig.~\ref{Stationary_distribution_different_lambda}~(a) and different mean-reverting rates $\lambda$, see Fig.~\ref{Stationary_distribution_different_lambda}~(b). The position distribution at any time $\tau \Delta t$ is roughly approximated from a histogram of an ensemble of $N=10^4$ particles. In particular, $P(x)\approx hist(bin(x))/\sum_{bin}hist(bin)$, where $bin(x)$ is the bin containing a specific position $x$ and $hist(y)$ is the number of particles in the $y$-th bin. As a convention, in $bin(x)$, we calculate the average number of particle positions between two successive time steps. 

As it follows from the numerical results and confirmed by the simulations, the two cases can be distinguished. In the first case, when $r\geq\lambda$, it is evident that there is a singular point with a peak at $x=x_0$, and as the value of the mean-reverting coefficient $\lambda$ increases, the stationary probability distribution around the long-term mean value $\mu$ increases as well. In the second case when $r<\lambda$, the singular point at the reset point $x=x_0$ appears again, but now the peak of the function  has moved away from the reset point, and is around the long-term mean value $\mu$.  Note also that the greater the coefficient $\lambda$ is, the closer the peak is to the point $x=\mu$. 

\subsection{Mean squared displacement}

The MSD can be defined from the renewal Eq.~(\ref{pdf renewal ou}), as well. It reads
\begin{align}\label{msd_with_resetting_renewal}
    \langle x^2(t)\rangle_{r}=e^{-rt}\,\langle x^2(t)\rangle +\int_{0}^{t}r\,e^{-rt'}\,\langle x^2(t')\rangle\,dt',
\end{align}
where $\langle x^2(t)\rangle$ is the MSD without resetting~(\ref{msd standard ou mu}). Performing the Laplace transform of Eq.~(\ref{msd_with_resetting_renewal}) and then after small algebra and the inverse Laplace transform, we obtain
\begin{align}\label{msd_with_resetting}
    \langle x^2(t)\rangle_{r}
    &=\frac{\sigma^2+r\,x_0^2}{r+2\lambda}+\left(x_0^2-\frac{\sigma^2}{2\lambda}\right)\frac{2\lambda}{r+2\lambda}\,e^{-(r+2\lambda)t}\nonumber\\&+\frac{2\lambda(\lambda \mu^2+ \mu r x_0 )}{(r+\lambda)(r+2 \lambda)}\left[1-e^{-(r+2 \lambda)t}\right].
\end{align}

In the long time limit $(t\rightarrow\infty)$, the MSD~(\ref{msd_with_resetting}) reads
\begin{align}
     \langle x^2(t)\rangle_{r} \sim \frac{\sigma^2+r\,x_0^2}{r+2\lambda}+\frac{2\lambda(\lambda \mu^2+ \mu r x_0 )}{(r+\lambda)(r+2 \lambda)},
\end{align}

For $\mu=0$, Eq.~(\ref{msd_with_resetting}) turns to
\begin{align}
    \langle x^2(t)\rangle_{r}
    =\frac{\sigma^2+r\,x_0^2}{r+2\lambda}+\left(x_0^2-\frac{\sigma^2}{2\lambda}\right)\frac{2\lambda}{r+2\lambda}\,e^{-(r+2\lambda)t},
\end{align}
and the long-time limit yields $$\langle x^2(t)\rangle_{r}\sim\frac{\sigma^2+r\,x_0^2}{r+2\lambda},$$which for $r=0$ recovers the O-U result without resetting, $$\langle x^2(t)\rangle_{r=0}=\frac{\sigma^2}{2\lambda}+\left(x_0^2-\frac{\sigma^2}{2\lambda}\right)e^{-2\lambda t}.$$

\begin{widetext}




\begin{figure}[h!]
\centering
\centerline{\includegraphics[width=15cm]{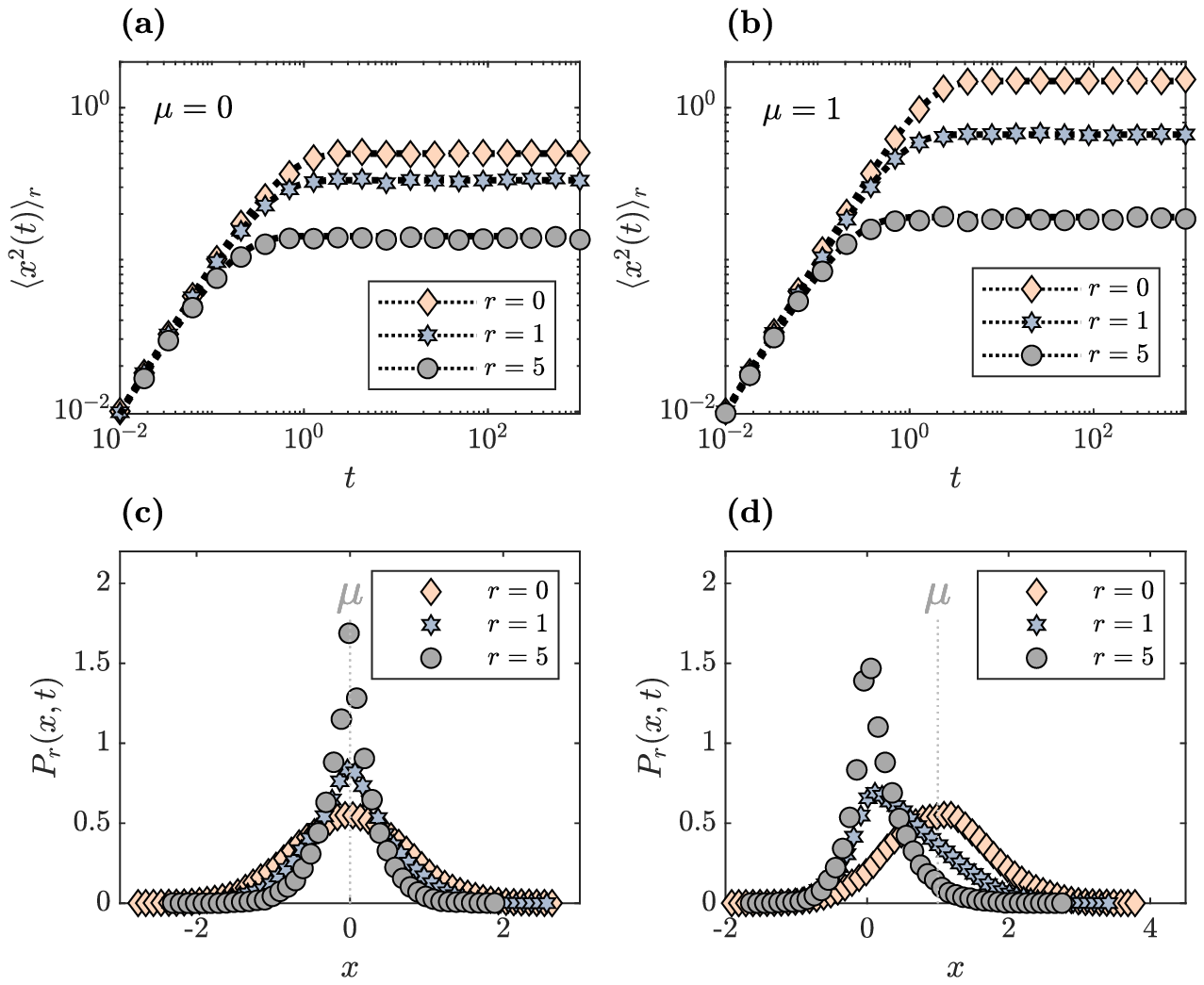}}
\caption{\textbf{(a)} MSD as a function of time for the O-U process with different resetting rates and $\mu=0$; \textbf{(b)} Same as (a) with $\mu=1$; \textbf{(c)} Long time behavior of the PDF with different resetting rates and $\mu=0$; \textbf{(d)} Same as (c) with $\mu=1$. We use $\sigma^2=1$, $\lambda=1$, $T=1000$, $dt=0.01$, and an initial ensemble of $N=10^4$ trajectories. For the Monte-Carlo simulations, the Langevin  Eq.~(\ref{langevin_resetting}) is used, and Eq.~(\ref{msd_with_resetting}) for the analytical solution of the MSD (dashed lines in \textbf{(a)} and \textbf{(b)}). Here the analytical solution of the MSD is being used to acquire precise simulation parameters, needed for plotting the PDF.   
 }\label{msd-ou-comparison2}
\end{figure}



\end{widetext}

The MSDs~(\ref{msd_with_resetting}) for different resetting rates are depicted in Figs.~\ref{msd-ou-comparison2}~(a) and~(b) with $\mu=0$ and $\mu=1$ respectively, and the results are compared with those obtained by simulations. The MSD at some time $\tau \Delta t$, where $\tau$ is an integer and $\Delta t$ is a discrete time increment, is calculated as an ensemble average of $N=10^4$ particles. As it is seen from the numerical results, the influence of the long-term mean value $\mu$ on the MSD is straightforward: the large is $\mu$, the larger is the MSD. The long time behaviour of the PDFs are presnted in Figs.~\ref{msd-ou-comparison2}~(c) and~(d) for $\mu= 0$ and $\mu = 1$, respectively. Another important result relates to the resetting rate $r$. Namely, for the larger $r$, the probability to find the particle near the initial condition is larger, and correspondingly the smaller the MSD is. Correspondingly, for $r=0$, the maximum of the PDF is at $x\sim \mu$.

\section{Ornstein-Uhlenbeck process on comb}\label{OU-comb}

\begin{figure}
\centering
\includegraphics[width=7.5cm]{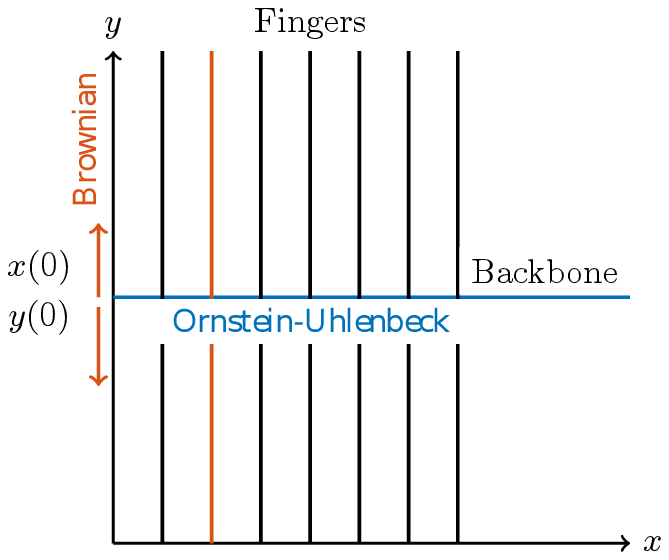}
\caption{Two dimensional comb structure. The backbone along the $x$ axis has continuously distributed fingers (or branches) along the $y$ axis. The O-U transport takes place along the backbone, while Brownian motion is in fingers.}\label{comb_model}
\end{figure}

In this section, we employ a comb model for the O-U process. We follow the phenomenological Fokker-Planck equation, suggested in Ref.~\cite{Arkhincheev1991} and extensively explored in a variety of applications, see Refs.~\cite{Iomin2018,book_ws}. According to the comb model, the two dimensional transport consists of two independent processes, shown in Fig.~\ref{comb_model}. The first is the O-U process, which takes place along the $x$ axis exactly at $y=0$, and this axis is called the backbone, and the corresponding motion is the backbone transport. In the $y$ direction, there is Brownian motion with the diffusion coefficient $\sigma^2_y/2$, which is the side-branched motion and the direction is called fingers, or side-branches.
The corresponding Fokker-Planck equation for this process is
\begin{align}\label{FPeq OUprocess comb} 
\frac{\partial}{\partial t}P(x,y,t)=\delta(y)\,L_{FP,x}P(x,y,t)+\frac{\sigma_y^2}{2}\,\frac{\partial
^2}{\partial y^2}P(x,y,t),
\end{align}
with the initial condition $P(x,y,t=0)=\delta(x-x_0)\,\delta(y)$ and zero boundary conditions at infinity. The Fokker-Planck operator on the backbone reads\footnote{Note that the complete form of the Fokker-Planck operator contains $\delta(y)$, as well. We however keep its present form 
of Eq.~(\ref{FP_operator}) to separate the  O-U process and to stress it in the ensuing analysis.}
\begin{align}\label{FPx_operator}
    L_{FP,x}\equiv\lambda\,\frac{\partial}{\partial x}(x-\mu)+\frac{\sigma_x^2}{2}\,\frac{\partial
^2}{\partial x^2}.
\end{align}
Note that $\lambda\delta(y)$ now is  the rate of mean-reversion, such that $\lambda$ is the  velocity of mean-reversion. Bearing this change in mind, we also admit that the backbone diffusion coefficient is $\delta(y)\sigma_x^2/2$. In what follows we keep the same notation for the velocity of mean-reversion to stress the O-U process along the backbone.


\subsection{Fokker-Planck equations for the marginal probability density functions}

Inferring the differential equations for the movement of the particle along the backbone and the fingers separately, we introduce the corresponding marginal PDFs as follows
\begin{align}\label{marginal pdf x}
p_1(x,t)=\int_{-\infty}^{\infty}P(x,y,t)\,dy,
\end{align}
and 
\begin{align}\label{marginal pdf y}
p_2(y,t)=\int_{-\infty}^{\infty}P(x,y,t)\,dx.    
\end{align}
Introducing new notations\footnote{Note also that the diffusion coefficient in the fingers is $\mathcal{D}_y=\sigma_y^2/2$, while the backbone diffusion coefficient is $\delta(y)\,\sigma_x^2/2=\delta(y)\,\mathcal{D}_x$.} $\mathcal{D}_x = \frac{\sigma_x^2}{2}$, $\mathcal{D}_y = \frac{\sigma_y^2}{2}$, in Eqs.~(\ref{FPeq OUprocess comb}) and~(\ref{FPx_operator}), we arrive at the comb equation as follows
\begin{align}\label{FPeq OUprocess comb app2} 
\frac{\partial}{\partial t}P(x,y,t)=\delta(y)\,L_{FP,x}P(x,y,t)+\mathcal{D}_y\,\frac{\partial
^2}{\partial y^2}P(x,y,t).
\end{align}
The Laplace transform of Eq.~(\ref{FPeq OUprocess comb app2}) yields
\begin{align}\label{diff_eq_laplace}
s\,\hat{P}(x,y,s) - \delta(x-x_0)\,\delta(y) &= \delta (y)\,L_{FP,x}\hat{P}(x, y, s)\nonumber\\& + \mathcal{D}_y\, \frac{\partial^2}{\partial y^2} \hat{P}(x, y, s).
\end{align}
The solution $\hat{P}(x,y,s)$ is presented in the following form
\begin{align}\label{P(x,y,s)}
\hat{P}(x, y, s) = \hat{g}(x, s)\, e^{-\hat{r}(x, s)\,|y|},
\end{align}
that yields the backbone marginal PDF as follows\footnote{Do not confuse $\hat{r}(x,s)$
with the reset rate $r$.}
\begin{align}\label{p1 g r}
    \hat{p}_{1}(x,s)
    =\int_{-\infty}^{\infty}\hat{g}(x,s)\,e^{-\hat{r}(x,s)|y|}\,dy=\frac{2\hat{g}(x,s)}{\hat{r}(x,s)}.
\end{align}
Taking into account the expression for the step sign function $\frac{d}{dy}|y|=2\theta(y)-1$, where $\theta(y)$ is the Heaviside theta function, one obtains
\begin{align}
\frac{\partial}{\partial y}\hat{P}(x,y,s) = -\hat{g}(x,s)\,\hat{r}(x,s)\,e^{-\hat{r}(x,s)|y|}\big[2\theta(y)-1\big]. 
\end{align}
By using the property of the Heaviside function: $\frac{d}{dy}\theta(y)=\delta(y)$, we obtain
\begin{align}\label{P_y^2}
    \frac{\partial^2}{\partial y^2} \hat{P}(x, y, s) &= -\hat{r}(x, s)\,\hat{g}(x, s)\,\big\{ 2\delta(y)\nonumber\\& - \hat{r}(x, s) \left[ 2\theta (y) - 1\right]^{2}\big\} \, e^{-\hat{r}(x, s)|y|}.
\end{align}
Taking into account Eqs.~(\ref{P(x,y,s)}) and (\ref{P_y^2}) and using the property $f(y)\,\delta(y)=f(0)\,\delta(y)$, we obtain Eq.~(\ref{diff_eq_laplace}) as follows
\begin{align}
    & s\,\hat{g}(x, s)\,e^{-\hat{r}(x, s)|y|} - \delta(x - x_0)\, \delta (y) = \lambda\, \delta (y)\, \hat{g}(x,s)\nonumber\\&+\lambda\, \delta(y)\, (x-\mu)\,\frac{\partial}{\partial x}\hat{g}(x,s) +\mathcal{D}_x\, \delta(y)\, \frac{\partial^2}{\partial x^2}\hat{g}(x,s)\nonumber \\& +\mathcal{D}_y\, \hat{r}^2(x,s)\,\hat{g}(x,s)\,e^{-\hat{r}(x,s)|y|}-2\delta(y)\,\mathcal{D}_y\, \hat{r}(x,s)\,\hat{g}(x,s).
\end{align}    
Thus, we arrive at the system of two equations
\begin{equation}\label{r_x_s}
s=\mathcal{D}_{y}\,\hat{r}^2(x,s) \quad \rightarrow \quad \hat{r}(x,s)=\sqrt{\frac{s}{\mathcal{D}_y}},
\end{equation}
and
\begin{align}\label{equ_g_x_s}
-\delta(x-x_0)&=\lambda\, \hat{g}(x,s)+\lambda\, (x-\mu)\, \frac{\partial}{\partial x}\hat{g}(x,s) \nonumber\\&+\mathcal{D}_{x}\,\frac{\partial^2}{\partial x^2}\hat{g}(x,s)-2\mathcal{D}_{y}\,\hat{r}(x,s)\,\hat{g}(x,s).
\end{align}
From Eqs.~(\ref{p1 g r}) and (\ref{r_x_s}) we derive
\begin{equation}
    \hat{g}(x,s)=\frac{1}{2}\sqrt{\frac{s}{\mathcal{D}_y}}\,\hat{p}_1(x,s).
\end{equation}
Now by substituting for $\hat{g}(x,s)$ in Eq.~(\ref{equ_g_x_s}) we get
\begin{align}\label{fractional_eq_in_laplace}
s^{1/2}\, \hat{p}_1(x, s) &- s^{-1/2}\,\delta(x - x_0) =
\frac{\lambda}{2 \sqrt{{\mathcal{D}_y}}}\, \hat{p}_1(x, s)\nonumber\\& +
\frac{\lambda\, (x - \mu)}{2 \sqrt{\mathcal{D}_y}}\,  \frac{\partial}{\partial x}\, \hat{p}_1 (x, s) +
\frac{\mathcal{D}_x}{2 \sqrt{{\mathcal{D}_y}}}\,\frac{\partial^2}{\partial x^2}\hat{p}_1 (x, s), 
\end{align}
which by the inverse Laplace transform yields the following time fractional diffusion equation
\begin{align}\label{FPeq OUprocess comb x} 
\frac{\partial}{\partial t}p_1(x,t)=\frac{1}{2\sqrt{\mathcal{D}_y}}\,{_{\text{RL}}}D_{t}^{1/2}L_{FP,x}p_1(x,t),
\end{align}
where ${_{\text{RL}}}D_{t}^{\mu}$ is the Riemann-Liouville fractional derivative~(\ref{rl derivative}) of order $\mu=1/2$.
It can be also written in terms of the Caputo fractional derivative, as follows
\begin{align}\label{FPeq OUprocess comb x caputo} 
{_{\text{C}}}D_{t}^{1/2}p_1(x,t)=\frac{1}{2\sqrt{\mathcal{D}_y}}\,L_{FP,x}\,p_1(x,t),
\end{align}
where ${_{\text{C}}}D_{t}^{\mu}$ is the Caputo fractional derivative~(\ref{caputo derivative}) of order $\mu=1/2$. Analytical properties of the fractional O-U process, described by the 
fractional Fokker-Planck equation~(\ref{FPeq OUprocess comb x}) has been discussed in great detail in Ref.~\cite{MeKl2000}. 

Integrating Eq.~(\ref{FPeq OUprocess comb}) with respect to $x$, we obtain the Fokker-Planck equation for the marginal PDF along fingers, which reads
\begin{align}\label{FPeq OUprocess2} 
\frac{\partial}{\partial t} p_2(y,t)=\mathcal{D}_y\,\frac{\partial^2}{\partial y^2}p_2(y,t).
\end{align}
The solution to this equation is the Gaussian PDF, as expected, since the particle performs Brownian motion along the fingers. 

\subsection{First moment and mean squared displacement}

The MSD along the backbone can be found by multiplying both sides of the Eq.~(\ref{FPeq OUprocess comb x caputo}) with $x^2$ and integrating with respect to $x$,
\begin{align}\label{msd OUprocess comb x1} 
{_{\text{C}}}D_{t}^{1/2}\langle x^2(t)\rangle_{\text{c}}&=\frac{\lambda}{2\sqrt{\mathcal{D}_y}}\int_{-\infty}^{\infty}x^2\frac{\partial}{\partial x}\big[(x-\mu)\,p_1(x,t)\big]dx\nonumber\\&+\frac{\mathcal{D}_x}{2\sqrt{\mathcal{D}_y}}\int_{-\infty}^{\infty}x^2\frac{\partial^2}{\partial x^2}p_1(x,t)\,dx
\end{align}
that yields
\begin{align}\label{msd OUprocess comb x} 
{_{\text{C}}}D_{t}^{1/2}\langle x^2(t)\rangle_{\text{c}}&=-\frac{\lambda}{\sqrt{\mathcal{D}_y}}\,\langle x^2(t)\rangle_{\text{c}}\nonumber\\&+\frac{\lambda \mu}{2 \sqrt{\mathcal{D}_y}}\,\langle x(t)\rangle_{\text{c}}+\frac{\mathcal{D}_x}{\sqrt{\mathcal{D}_y}}.
\end{align}

Equation for the mean value $\langle x(t) \rangle$ is obtained in the same way, and it reads
\begin{align}
  {_{\text{C}}}D_{t}^{1/2}\langle x(t)\rangle_{\text{c}}=-\frac{\lambda}{2 \sqrt{\mathcal{D}_y}}\,\langle x(t)\rangle_{\text{c}}+\frac{\lambda \mu}{2 \sqrt{\mathcal{D}_y}}.
\end{align}
In Laplace space, the mean value is
\begin{align}\label{mean comb laplace}
     \langle \hat{x}(s)\rangle_{\text{c}}=\frac{x_0\, s^{-1/2}}{s^{1/2}+\frac{\lambda}{2\sqrt{\mathcal{D}_y}}}+\frac{\lambda \mu}{2 \sqrt{\mathcal{D}_y}}\,\frac{s^{-1}}{s^{1/2}+\frac{\lambda}{2\sqrt{\mathcal{D}_y}}}.
\end{align}
Performing the inverse Laplace transform, we get
\begin{align}\label{mean comb ou}
    \langle x(t)\rangle_{\text{c}}&= x_0\, E_{1/2} \left(-\frac{\lambda}{2\sqrt{\mathcal{D}_y}}t^{1/2} \right)\nonumber\\&+\frac{\lambda \mu}{2 \sqrt{\mathcal{D}_y}}\,t^{1/2}\,E_{1/2, 3/2} \left(-\frac{\lambda}{2\sqrt{\mathcal{D}_y}}t^{1/2} \right),
\end{align}
where $E_{\alpha}(z)$ and $E_{\alpha,\beta}(z)$ are the one and two parameter Mittag-Leffler functions, respectively, see Eqs.~(\ref{one parameter ML}) and~(\ref{two parameter ML}) in Appendix~\ref{appB}.

Now the exact expression for the MSD in Eq.~(\ref{msd OUprocess comb x}) can be obtained. Performing the Laplace transform of Eq.~(\ref{msd OUprocess comb x}) and taking into account 
Eq.~(\ref{mean comb laplace}), we obtain the Laplace image of the MSD as follows
\begin{align}\label{msd OUprocess comb x laplace2} 
\langle \hat{x}^2(s)\rangle_{\text{c}}&=x_0^2\,\frac{s^{-1/2}}{s^{1/2}+\frac{\lambda}{\sqrt{\mathcal{D}_y}}}+\frac{\mathcal{D}_x}{\sqrt{\mathcal{D}_y}}\,\frac{s^{-1}}{s^{1/2}+\frac{\lambda}{\sqrt{\mathcal{D}_y}}}\nonumber\\&+
\frac{\lambda \mu x_0}{\sqrt{\mathcal{D}_y}}\, \frac{s^{-1/2}}{\left(s^{1/2}+\frac{\lambda}{\sqrt{\mathcal{D}_y}}\right)\left(s^{1/2}+\frac{\lambda}{2\sqrt{\mathcal{D}_y}}\right)}\nonumber\\&+\frac{\lambda^2 \mu^2}{2 \sqrt{\mathcal{D}_y}}\,\frac{s^{-1}}{\left(s^{1/2}+\frac{\lambda}{\sqrt{D_y}}\right)\left(s^{1/2}+\frac{\lambda}{2\sqrt{\mathcal{D}_y}}\right)}.
\end{align}
The inverse Laplace transform yields the expression for the MSD along the backbone,
\begin{align}\label{MSD_comb_mu}
       & \langle x^2(t)\rangle_{\text{c}}=x_0^2\,E_{1/2}\left(-\frac{\lambda}{\sqrt{\mathcal{D}_y}}\,t^{1/2}\right)\nonumber\\&+\frac{\mathcal{D}_x}{\sqrt{\mathcal{D}_y}}\,t^{1/2}\,E_{1/2,3/2}\left(-\frac{\lambda}{\sqrt{\mathcal{D}_y}}\,t^{1/2}\right)
    \nonumber \\&+\frac{\lambda \mu x_0}{2 \sqrt{\mathcal{D}_y}}\,t^{1/2}\,E_{(1/2,1),3/2}\left(-\frac{3\lambda}{2\sqrt{\mathcal{D}_y}}\,t^{1/2}\,\frac{\lambda^2}{2\mathcal{D}_y}\,t\right)\nonumber \\&+\frac{\lambda^2 \mu^2}{4\mathcal{D}_y} \, t \, E_{(1/2,1),2}\left(-\frac{3\lambda}{2\sqrt{\mathcal{D}_y}}\,t^{1/2},-\frac{\lambda^2}{2\mathcal{D}_y}\,t\right)
    ,
\end{align}
where 
$E_{(\alpha_1,\alpha_2),\beta}(z;\lambda_1,\lambda_2)$ is the multinomial Mittag-Leffler function, see Eq.~(\ref{multinomial ml}). The long time limit yields the saturation behaviour of the MSD
\begin{align}\label{msd OUprocess comb x laplace2 long time} 
\langle x^2(t)\rangle_{\text{c}}\sim\mu^2 \sqrt{\mathcal{D}_y}+\frac{\mathcal{D}_x}{\lambda}.
\end{align}
However, the transition to the constant MSD is slower (of the power-law decay) than the one for the one dimensional O-U process (of the exponential decay) due to the fact that the particle is hindered in the fingers before it turns back to the backbone transport. The power-law decay to the constant value can be shown by asymptotic analysis of the exact MSD~(\ref{msd OUprocess comb x laplace2 long time}). For $\lambda=0$ we recover the result for the comb model, $\langle x^2(t)\rangle_{\text{c}}=x_0^2+\frac{\mathcal{D}_x}{\sqrt{\mathcal{D}_y}}\frac{t^{1/2}}{\Gamma(3/2)}$, as expected. 

\section{Ornstein-Uhlenbeck process on comb with resetting}\label{OU-comb-resetting}

In this section we extend the problem of the O-U process on a comb by introducing stochastic resetting~\cite{prr}. We consider resets to the initial position $(x,y)=(x_0,0)$ with the resetting rate $r$. This results in the following Fokker-Planck equation
\begin{align}\label{FPeq OUprocess comb reset} 
\frac{\partial}{\partial t}P_r(x,y,t)&=\delta(y)\,L_{FP,x}P_r(x,y,t)+\frac{\sigma_y^2}{2}\,\frac{\partial
^2}{\partial y^2}P_r(x,y,t)\nonumber\\&-r\,P_r(x,y,t)+r\, \delta(x-x_0)\,\delta(y)
\end{align}
with the initial condition $P_r(x,y,t=0)= \delta (x-x_0)\,\delta(y)$ and zero boundary conditions at infinity. We analyze the transport properties of the particle on the backbone and inside the fingers separately, that is, we calculate the marginal PDFs $p_1(x,t)$ and $p_2 (y,t)$.

\subsection{Numerical simulations: Coupled Langevin equations}\label{Langevin_eq_for_comb_with_reset}

The motion with resetting on the two dimensional comb structure can be simulated by the following coupled Langevin equations~\cite{prr,mathematics} (in case of no resetting we refer to~\cite{mendez1,lenzi1})
\begin{align}\label{x-axis langevin reset}
\left\lbrace\begin{array}{l l l}
     x(\tau\Delta t) = x[(\tau-1)\Delta t]+ A(y)\big(\lambda \left[\mu-x((\tau-1)\Delta t)\right]\big) \\
     +\sqrt{2D_xA(y)\Delta t}\,\xi_x[(\tau-1)\Delta t], \,\, \text{with prob.} \ (1-r\Delta t), \\ \\ x(\tau\Delta t)=x(0), \,\, \text{with  prob.} \ r\Delta t,
\end{array}\right.
\end{align}
for the movement along the backbone, and
\begin{align}\label{y-axis langevin reset}
\left\lbrace\begin{array}{l l l}
     y(\tau\Delta t) = y[(\tau-1)\Delta t]+\sqrt{2D_y\,\Delta t}\,\xi_y[(\tau-1)\Delta t], \\ \text{with prob.} \ (1-r\Delta t), \\ \\
     y(\tau\Delta t)=y(0), \,\, \text{with prob.} \ r\Delta t,
\end{array}\right.
\end{align}
for the Brownian motion along the fingers. Here $\xi_i$, $i=\{x,y\}$, is the same white noise as in Eq.~(\ref{langevin_resetting}) with zero mean, $\langle \xi_i(\tau \Delta t) \rangle = 0$, and correlation function $\langle \xi_i(\tau \Delta t)\xi_i(\tau'\Delta t) \rangle = \delta((\tau-\tau')\Delta t)$. The function $A(y)$ is introduced to describe the motion along the backbone at $y=0$, where $A(y)$ is the approximation of the Dirac $\delta$ function by means of the expression $A(y)=\frac{1}{\sqrt{2\pi}\sigma_\delta}\exp\left(-y^2/(2\sigma_\delta^2)\right)$, $\sigma _{\delta}\rightarrow0$. Here $\sigma_{\delta}$ is taken such that it must be of order of $2\sigma_{\delta} \gtrsim \sqrt{2\mathcal{D}_y\Delta t}$. We have found that if we take the value of $\sigma_{\delta}$ to be $\sigma_{\delta} =\frac{\sqrt{2\mathcal{D}_y\Delta t}}{2}+\varepsilon$ where $\varepsilon=10^{-3}$, more than satisfactory matching of the analytical and simulated results are obtained. For the simulations of the marginal PDF along the backbone, the diffusion coefficient along the backbone and the mean-reverting rate are renormalized by factor $1/[2\sqrt{\mathcal{D}_y}]$, see Refs.~\cite{prr,mathematics} and Eq.~(\ref{fpe p1 comb}).


Results of the simulated trajectories according to the coupled Langevin Eqs.~(\ref{x-axis langevin reset}) and~(\ref{y-axis langevin reset}) with resetting to $x=x_0=0$ are presented in Fig.~\ref{trajectories_comb}. The plateaus with the fixed $x$ in the backbone dynamics reflect the waiting times due to diffusion in the fingers.


\subsection{Fokker-Planck equations for the marginal probability density functions}

We find the differential equations for the marginal PDFs along the backbone and fingers, with the same procedure as in the case of diffusion on the comb model without resetting. Thus,  integration of Eq.~(\ref{FPeq OUprocess comb reset}) with respect to $x$ yields
\begin{align}\label{fpe p2 comb}
    \frac{\partial}{\partial t}p_{2,r} (y,t)=\mathcal{D}_y \,\frac{d}{d t}\int_{0}^{t}e^{-r(t-t')}\,
      \frac{\partial^2}{\partial y^2}p_{2,r}(y,t')\,dt'.
\end{align}
It describes Brownian motion with resetting along the fingers. 

Performing integration with respect to $y$, we obtain 
\begin{align}\label{fpe p1 comb}
    \frac{\partial}{\partial t}p_{1,r} (x,t)=\frac{1}{2 \sqrt{\mathcal{D}_y}}\,{_{\textrm{TRL}}}D_{0+}^{1/2}\,L_{FP,x}p_{1,r} (x,t),
\end{align}
which is the equation for the transport along the backbone, where $_{\textrm{TRL}}D_{0+}^{\mu}f(t)$ is the so-called {\it tempered} Riemann-Liouville fractional derivative~(\ref{temperedRLderivative}) of order $\mu=1/2$ with tempering parameter $r$. Again, here we use $\sigma^2_i/2=\mathcal{D}_i$, for $i=\{x,y\}$. From the subordination approach, it can be shown that the marginal PDF along the backbone can be obtained from the PDF of the standard O-U process or the PDF for the comb without resetting, i.e.,
\begin{align}\label{p1_r-P_o}
    \hat{p}_{1,r} (x,s)&=\frac{1}{s\, \hat{\eta}_r (s)}\, \hat{P}_0\left(x,1/\hat{\eta}_r (s)\right)\nonumber\\&=\frac{(s+r)^{1/2}}{s}\,\hat{P}_{0}(x,(s+r)^{1/2})\nonumber\\&=\frac{s+r}{s}\,\hat{p}_{1}(x,s+r),
\end{align}
which actually is the renewal equation for the marginal PDF, i.e., see also~\cite{mathematics},
\begin{align}
    p_{1,r}(x,t)=e^{-rt}\,p_1(x,t)+\int_{0}^{t}r\,e^{-rt'}\,p_{1}(x,t')\,dt'.
\end{align}

Comparing the results for the PDF for the O-U process without resetting, presented in Fig.~\ref{PDF_no_comb_plot}~(a)
with the same results for the comb structure presented in Fig.~\ref{comb-plot3}~(a), it becomes evident the staggering of the particles diffusing on the backbone as the result of their getting stuck in the fingers. This is mostly visible for the PDF at $t=1$. For example in Fig.~\ref{comb-plot3}~(a), there is a finite probability to find the particle near the initial position $x_0=0$, while it is not the case in Fig.~\ref{PDF_no_comb_plot}~(a), where this probability is less dispersed and the particles are concentrated around some point in the temporal evolution of the process. The corresponding  cases with resetting are compared in Figs.~\ref{PDF_no_comb_plot}~(b) and~\ref{comb-plot3}~(b). As shown in Fig.~\ref{PDF_no_comb_plot}~(b) the one dimensional O-U process with resetting tends to the homogeneous distribution of particles in the interval $x\in (x_0,\mu)$. This situation changes drastically in the comb geometry, shown in Fig.~\ref{comb-plot3}~(b), where the asymptotic marginal PDF has a well defined maximum. Obviously, this shape of the PDF results from the long time trapping of the particles inside fingers. Figs.~\ref{comb-plot3}~(c) and~(d) are the evidence of another property of hindering of relaxation due to the parameters $\lambda$ and $r$. As it follows from the numerical results, the larger values of the mean reverting velocity lead to stronger localization of the initial distribution. Resetting is responsible for the  decreasing of the relaxation rate. Comparing Figs.~\ref{comb-plot3}~(c) and~\ref{PDF_no_comb_plot}~(c) for $\lambda=0$, the comb geometry effect becomes evident, which however is attenuated by the O-U process for $\lambda\neq 0$.


\begin{widetext}

\begin{figure}[h!]
\centering
\includegraphics[width=17cm]{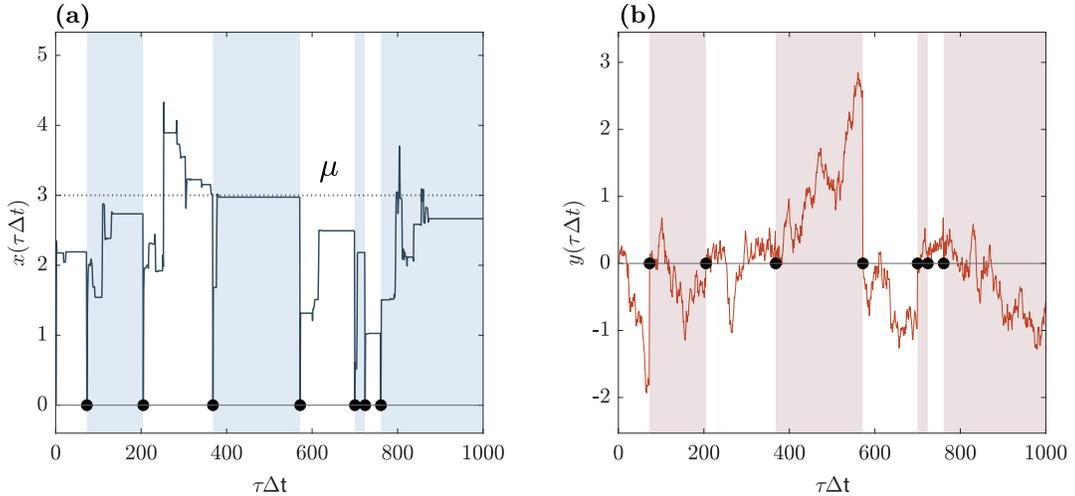}
\caption{Trajectories of the particle on the backbone (a) and in the fingers (b) according to the Langevin equations~(\ref{x-axis langevin reset}) and (\ref{y-axis langevin reset}) for $x_0=0$, $\mu = 3$, $\lambda=3$, $\sigma_x=\sigma_y=1$, $r=0.8$, $\Delta t=0.01$. The trapping of the tracer in the fingers is reflected by plateaus of the backbone's trajectory.}\label{trajectories_comb}
\end{figure}

\begin{figure}
\centering
\includegraphics[width=17cm]{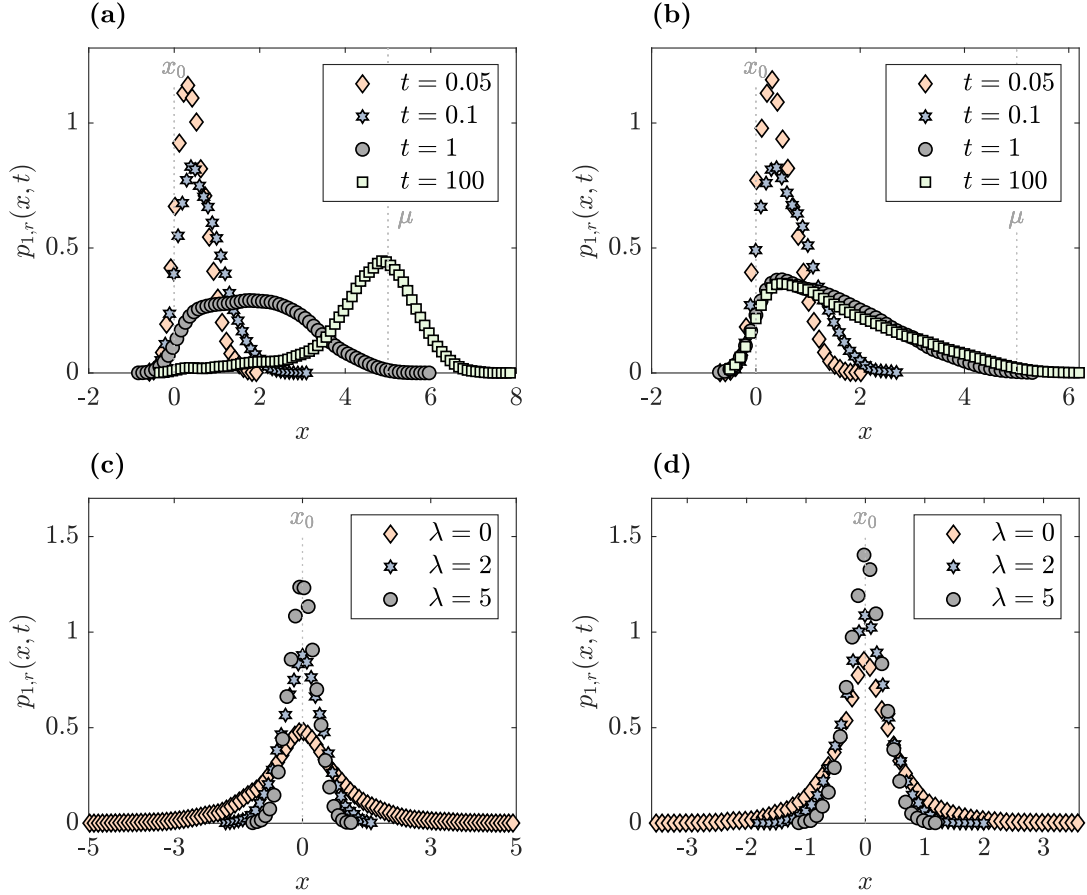}
\caption{Simulations of the marginal PDF along the backbone, according to the Langevin equation ~(\ref{x-axis langevin reset}), using the renormalized diffusion coefficient and the mean-reverting velocity by the parameter 
$1/[2\sqrt{\mathcal{D}_y}]$, see Eq.~(\ref{fpe p1 comb}) and Refs.~\cite{prr,mathematics}: \textbf{(a)} Evolution of the PDF in time for $x_0=0$, $r=0$, $\mu = 5$, $\lambda=1$, $\sigma_x=\sigma_y=1$, $\Delta t=0.01$, and an initial ensemble of $N=10^4$ particle trajectories; \textbf{(b)} Same as (a) for the resetting rate $r=1$; \textbf{(c)} PDF for different values of $\lambda$ and $x_0=0$, $t=5$, $\mu = 0$, $\sigma_x=\sigma_y=1$, $\Delta t=0.01$, and an initial ensemble of $N=10^4$ trajectories without resetting; \textbf{(d)} Same as (c) for the resetting rate $r=1$. }\label{comb-plot3}
\end{figure}

\end{widetext}

    

From Eq.~(\ref{p1_r-P_o}) we find that in the long time limit the system approaches a NESS given by
\begin{align}\label{ness_comb}
    p_{1,r}^{st}(x)&=\lim_{t\rightarrow\infty}p_{1,r}(x,t)\nonumber\\&=\lim_{s\rightarrow0}s\,\hat{p}_{1,r}(x,s)=r\,\hat{p}_{1}(x,r).
\end{align}
In Fig.~\ref{msd-ou-comb-comparison2}~(c) and~(d) we present the marginal NESS in Eq.~(\ref{ness_comb}), obtained by numerical simulations.

\subsection{Mean squared displacement}

\begin{widetext}

\begin{figure}[h!]
\centering
\centerline{\includegraphics[width=15cm]{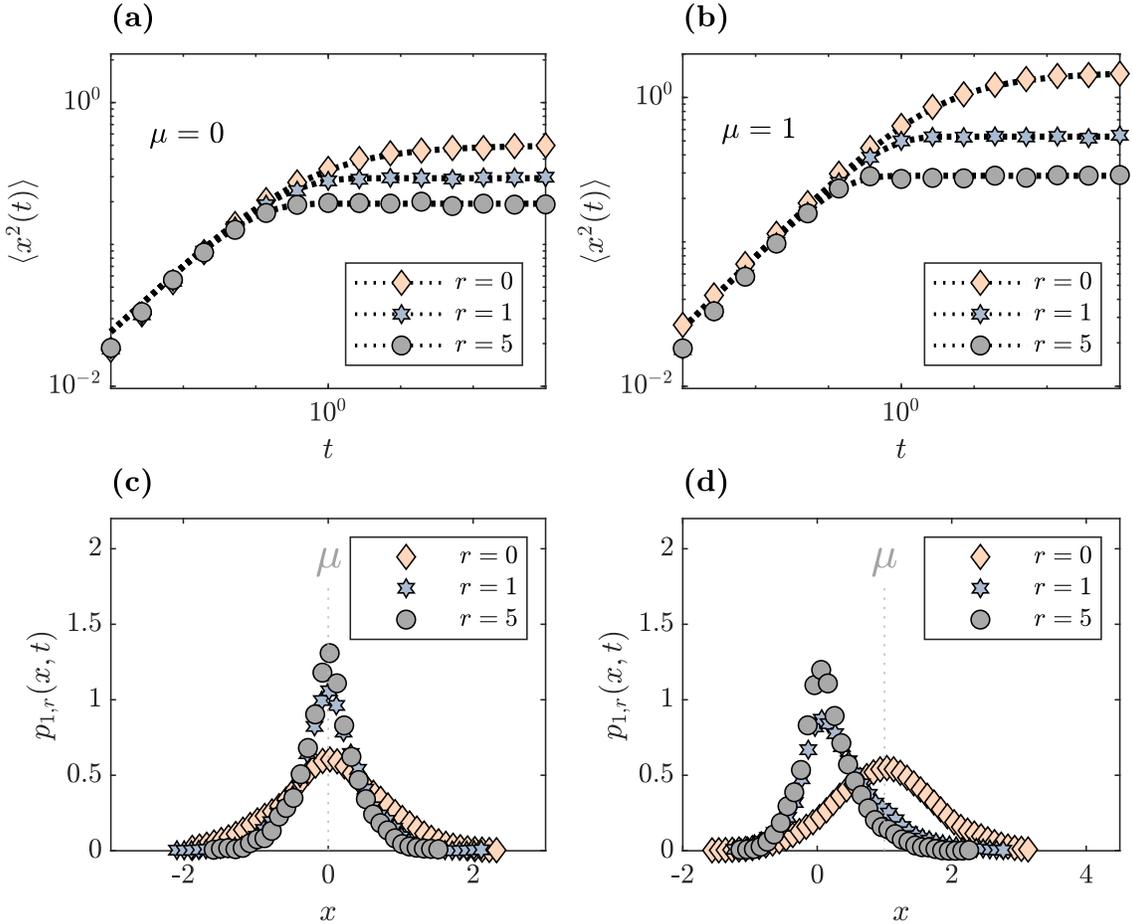}}
\caption{\textbf{(a)} The MSD for the O-U process on the backbone with different resetting rates; \textbf{(b)} Same as (a) for $\mu=1$; \textbf{(c)} Long time PDF with different reset rates for $\mu=0$, using the renormalized diffusion coefficients and mean-reverting rate, see Refs.~\cite{prr,mathematics}; \textbf{(d)} Same as (c) with $\mu=1$. We use $x_0=0$, $\lambda=1$, $\sigma_x=\sigma_y=1$, $dt=0.001$, $\sigma_{\delta}=0.016$, and an initial ensemble of $N=10^4$ trajectories. The $\sigma_{\delta}=0.016$ is used inside the approximation of the $\delta$ function - $A(y)$ in ~(\ref{x-axis langevin reset}). The dashed lines in (a) and (b) are the analytical solution for the MSD~(\ref{msd_comb_reset_final2}). As in the case of Fig.~\ref{msd-ou-comparison2}, the analytical solution here is being used as a way to acquire the simulation parameters needed for creating the PDFs.}\label{msd-ou-comb-comparison2}
\end{figure}

\end{widetext}

The corresponding MSD for the O-U process with resetting on the comb can be found from Eq.~(\ref{p1_r-P_o}). Thus, we find
\begin{align}
    \langle \hat{x}^2(s)\rangle_{\text{c},r}=\frac{s+r}{s}\,\langle \hat{x}^2(s+r)\rangle_{\text{c}},
\end{align}
and the renewal equation reads
\begin{align}\label{msd_comb_reset_final2}
\langle x^2(t)\rangle_{\text{c},r}=e^{-rt}\,\langle x^2(t)\rangle_{\text{c}}+\int_{0}^{t}r\,e^{-rt'}\,\langle x^2(t')\rangle_{\text{c}}\,dt',
\end{align}
where $\langle x^2(t)\rangle_{\text{c}}$ is the MSD~(\ref{MSD_comb_mu}) in absence of resetting. From Eq.~(\ref{msd_comb_reset_final2}), the long time behavior of the MSD reads
\begin{align}
\lim_{t\rightarrow\infty}\langle x^2(t)\rangle_{\text{c},r}&=
    \int_{0}^{\infty}r\,e^{-rt'}\,\langle \hat{x}^2(t')\rangle_{\text{c}}\,dt'=r\,\langle \hat{x}^2(r)\rangle_{\text{c}},
\end{align}
which eventually yields
\begin{align}
\label{msd_reset_in_laplace_append}
\langle x^2(t)\rangle_{\text{c},r}&\sim \frac{x_0^2\,r^{1/2}+\frac{\mathcal{D}_x}{\sqrt{D_y}}}{r^{1/2}+\frac{\lambda}{\sqrt{D_y}}}\nonumber\\&+\frac{\frac{\lambda\mu}{\sqrt{D_y}}\left(\frac{\lambda \mu}{2 \sqrt{\mathcal{D}_y}}+x_0\,r^{1/2}\right)}{\left(r^{1/2}+\frac{\lambda}{\sqrt{D_y}}\right)\left(r^{1/2}+\frac{\lambda}{2\sqrt{D_y}}\right)}.
\end{align}
In Fig.~\ref{msd-ou-comb-comparison2}~(a) and (b) the graphical representation of the MSD~(\ref{msd_comb_reset_final2}) obtained analytically and by numerical simulations is plotted, where the saturation of the MSD  in the long time limit is according to Eq.~(\ref{msd_reset_in_laplace_append}).

As obtained in Eq.~(\ref{msd_reset_in_laplace_append}) the saturation value of the MSD is a function of the  mean-reversion velocity $\lambda$. Therefore, the extremum equation
\begin{align}\label{lambda_derivative}
    \frac{\partial}{\partial \lambda} \langle x^2(t)\rangle_{\text{c},r} = 0
\end{align}
determines $\lambda_{\min}(r)$ for which the MSD is minimal. Considering the long time MSD~(\ref{msd_reset_in_laplace_append}) vs $\lambda$, we arrive at the conclusion that there are specific values of $\lambda$ and $r$, which minimize the MSD. As it follows from Figs.~\ref{comb-plot3}~(c) and~(d) and Figs.~\ref{msd-ou-comb-comparison2}~(c) and~(d), the evolution of the marginal PDF $p_{1,r}(x,t)$ depends essentially on the parameters $\lambda$ and $r$. The same situation is for the MSD. Therefore, the minimal value of the MSD, determined by Eq.~(\ref{lambda_derivative}) defines also the stronger localization of the  marginal PDF due to the resetting.

\section{Ornstein-Uhlenbeck process on fractal grid}\label{OUfractal}

Further geometrical generalization, is the consideration of the O-U process on a fractal grid structure~\cite{tr1,tr2,tr3}, which contains infinitely-uncountable number of backbones inside a finite-width strip. The backbones are positioned at $y=l_j \in \mathcal{S}_{\nu}$, where $\mathcal{S}_\nu$ is a fractal set with the fractal dimension $\nu$. The corresponding Fokker-Planck equation reads
\begin{align}\label{FPeq OUprocess grid} 
\frac{\partial}{\partial t}P(x,y,t)&=\sum_{l_j\in\mathcal{S}_{\nu}}\delta(y-l_j)\,L_{FP,x}P(x,y,t)\nonumber\\&+\frac{\sigma_y^2}{2}\,\frac{\partial
^2}{\partial y^2}P(x,y,t).
\end{align}

The geometrical structure of the equation means that the O-U process described by the Fokker-Planck operator $L_{FP,x}$, takes place along the fractal structure 
of the backbones.

In the section, we study the anomalous O-U transport along the fractal backbone structure, which is described by the marginal PDF $p_1(x,t)$. Integrating Eq.~(\ref{FPeq OUprocess grid}) with respect to $y$, we obtain
\begin{align}\label{FPeq OUprocess grid2} 
\frac{\partial}{\partial t}p_1(x,t)&=\sum_{l_j\in\mathcal{S}_{\nu}}\Big\{\lambda\,\frac{\partial}{\partial x}\big[x\,P(x,y=l_j,t)\big]\nonumber\\&+\frac{\sigma_x^2}{2}\,\frac{\partial^2}{\partial x^2}P(x,y=l_j,t)\Big\}.
\end{align}
The Laplace transform of Eq.~(\ref{FPeq OUprocess grid2}) yields
\begin{align}\label{FPeq OUprocess grid2 laplace} 
s\,\hat{p}_1(x,s)-p_{1}(x,t=0)&=\sum_{l_j\in\mathcal{S}_{\nu}}\Big\{\lambda\,\frac{\partial}{\partial x}\left[x\,\hat{P}(x,y=l_j,s)\right]\nonumber\\&+\frac{\sigma_x^2}{2}\,\frac{\partial^2}{\partial x^2}\hat{P}(x,y=l_j,s)\Big\}.
\end{align}
In Laplace space, we look for the solution to Eq.~(\ref{FPeq OUprocess grid2 laplace}) in the form
\begin{align}\label{eq11}
    \hat{P}(x,y,s)=g(x,s)\,e^{-\sqrt{\frac{s}{\sigma_y^2/2}}|y|},
\end{align}
from where it follows
\begin{align}\label{12}
    \hat{P}(x,y=l_j,s)=\hat{g}(x,s)\,e^{-\sqrt{\frac{s}{\sigma_y^2/2}}|l_j|}.
\end{align}
From Eq.~(\ref{eq11}), we obtain the Laplace image of the marginal PDF as follows
\begin{align}\label{p1_g}
    \hat{p}_{1}(x,s)=2\,\hat{g}(x,s)\,\sqrt{\frac{\sigma_y^2/2}{s}}.
\end{align}
The summation in Eqs.~(\ref{FPeq OUprocess grid2}) and~(\ref{FPeq OUprocess grid2 laplace}) is performed over the fractal set $\mathcal{S}_{\nu}$, and it corresponds to integration over the fractal measure $\mu_\nu\sim l^{\nu}$, and thus $\sum_{l_j\in\mathcal{S}_\nu}\rightarrow\frac{l^{\nu-1}}{\Gamma(\nu)}$ is the fractal density, while $d\mu_\nu=\frac{1}{\Gamma(\nu)}l^{\nu-1}\,dl$, see Ref.~\cite{tr1}. Thus, by summation over the fractal set, we have
\begin{align}
    \sum_{l_j\in\mathcal{S}_{\nu}}\hat{P}(x,y,s)&=\hat{g}(x,s)\,\frac{1}{\Gamma(\nu)}\int_{0}^{\infty}e^{-\sqrt{\frac{s}{\sigma_y^2/2}}l}\,l^{\nu-1}\,dl\nonumber\\&=g(x,s)\left(\frac{\sigma_y^2/2}{s}\right)^{\nu/2}\nonumber\\&=\frac{1}{2\left(\sigma_y^2/2\right)^{\frac{1-\nu}{2}}}\,s^{\frac{1-\nu}{2}}\,\hat{p}_1(x,s),
\end{align}
where the last line is according to Eq.~(\ref{p1_g}). From Eq.~(\ref{FPeq OUprocess grid2 laplace}), we find
\begin{align}\label{FPeq OUprocess grid2 laplace final} 
&s^{\frac{1+\nu}{2}}\,\hat{p}_1(x,s)-s^{\frac{1+\nu}{2}-1}\,p_{1}(x,t=0)\nonumber\\&=\frac{1}{2\left(\sigma_y^2/2\right)^{\frac{1-\nu}{2}}}\left\{\lambda\,\frac{\partial}{\partial x}\left[x\,\hat{p}_1(x,s)\right]+\frac{\sigma_x^2}{2}\,\frac{\partial^2}{\partial x^2}\hat{p}_1(x,s)\right\}.
\end{align}
The inverse Laplace transform of Eq.~(\ref{FPeq OUprocess grid2 laplace final} ) yields
\begin{align}\label{FPeq OUprocess grid2 final} 
{_{\text{C}}}D_{t}^{\frac{1+\nu}{2}}p_1(x,t)&=\frac{1}{2\left(\sigma_y^2/2\right)^{\frac{1-\nu}{2}}}\nonumber\\&\times\left\{\lambda\,\frac{\partial}{\partial x}\left[x\,p_1(x,t)\right]+\frac{\sigma_x^2}{2}\,\frac{\partial^2}{\partial x^2}p_1(x,t)\right\},
\end{align}
where ${_{\text{C}}}D_{t}^{\beta}$ is the Caputo fractional derivative~(\ref{caputo derivative}) of order $\frac{1}{2}<\beta=\frac{1+\nu}{2}<1$ (since $0<\nu<1$). 

From here, we find the MSD by multiplying both sides of the equation by $x^2$ and integrating over the $x$,
\begin{align}\label{msd OUprocess fractal comb x} 
{_{\text{C}}}D_{t}^{\frac{1+\nu}{2}}\langle x^2(t)\rangle=-\frac{\lambda}{\left(\sigma_y^2/2\right)^{\frac{1-\nu}{2}}}\,\langle x^2(t)\rangle+\frac{\sigma_x^2}{2\left(\sigma_y^2/2\right)^{\frac{1-\nu}{2}}},
\end{align}
which yields
\begin{align}\label{msd fractal comb laplace}
    \langle \hat{x}^2(s)\rangle&=x_0^2\,\frac{s^{\frac{1+\nu}{2}-1}}{s^{\frac{1+\nu}{2}}+\frac{\lambda}{\left(\sigma_y^2/2\right)^{\frac{1-\nu}{2}}}}\nonumber\\&+\frac{\sigma_x^2}{2\left(\sigma_y^2/2\right)^{\frac{1-\nu}{2}}}\,\frac{s^{-1}}{s^{\frac{1+\nu}{2}}+\frac{\lambda}{\left(\sigma_y^2/2\right)^{\frac{1-\nu}{2}}}}.
\end{align}
By the inverse Laplace transform, we eventually obtain
\begin{align}\label{MSD grid ou}
    \langle x^2(t)\rangle&=x_0^2\,E_{\frac{1+\nu}{2}}\left(-\frac{\lambda}{\left(\sigma_y^2/2\right)^{\frac{1-\nu}{2}}}t^{\frac{1+\nu}{2}}\right)\nonumber\\&+\frac{\sigma_x^2}{2\left(\sigma_y^2/2\right)^{\frac{1-\nu}{2}}}\,t^{\frac{1+\nu}{2}}\,E_{\frac{1+\nu}{2},\frac{3+\nu}{2}}\left(-\frac{\lambda}{\left(\sigma_y^2/2\right)^{\frac{1-\nu}{2}}}t^{\frac{1+\nu}{2}}\right).
\end{align}
For the short time scale, we find
\begin{align}
    \langle x^2(t)\rangle
    &\sim x_0^2+\frac{\sigma_x^2-2\lambda\,x_0^2}{2\left(\sigma_y^2/2\right)^{\frac{1-\nu}{2}}}\,\frac{t^{\frac{1+\nu}{2}}}{\Gamma(\frac{3+\nu}{2})}\nonumber\\&+\frac{2\lambda^2x_0^2-\lambda\sigma_x^2}{2\left(\sigma_y^2/2\right)^{1-\nu}}\,\frac{t^{1+\nu}}{\Gamma(2+\nu)},
\end{align}
while the long time limit yields saturation of the MSD,
\begin{align}\label{msd long fractal}
    \langle x^2(t)\rangle&\sim x_0^2\,\frac{\left(\sigma_y^2/2\right)^{\frac{1-\nu}{2}}}{\lambda}\,\frac{t^{-\frac{1+\nu}{2}}}{\Gamma(1-\frac{1+\nu}{2})}\nonumber\\&+\frac{\sigma_x^2}{2\left(\sigma_y^2/2\right)^{\frac{1-\nu}{2}}}\,t^{\frac{1+\nu}{2}}\,\frac{\left(\sigma_y^2/2\right)^{\frac{1-\nu}{2}}}{\lambda}\,t^{-\frac{1+\nu}{2}}\sim \frac{\sigma_x^2}{2\lambda},
\end{align}
with the power-law decay ($t^{-\frac{1+\nu}{2}}$) to the stationary value.

The obtained result for the MSD can be easily generalized for the presence of resetting, by using the renewal equation approach. Thus, the MSD in Laplace space reads
\begin{align}
    \langle \hat{x}^2(s)\rangle_r&=\frac{s+r}{s}\,\langle \hat{x}^2(s+r)\rangle\nonumber\\&=\frac{s+r}{s}\left[x_0^2\,\frac{(s+r)^{\frac{1+\nu}{2}-1}}{(s+r)^{\frac{1+\nu}{2}}+\frac{\lambda}{\left(\sigma_y^2/2\right)^{\frac{1-\nu}{2}}}}\right.\nonumber\\&\left.+\frac{\sigma_x^2}{2\left(\sigma_y^2/2\right)^{\frac{1-\nu}{2}}}\,\frac{(s+r)^{-1}}{(s+r)^{\frac{1+\nu}{2}}+\frac{\lambda}{\left(\sigma_y^2/2\right)^{\frac{1-\nu}{2}}}}\right],
\end{align}
where $\langle \hat{x}^2(s)\rangle$ is the MSD~(\ref{msd fractal comb laplace}) without resetting.

In the short time limit ($s\rightarrow\infty$) the MSD turns to the one obtained in the case without resetting~(\ref{MSD grid ou}). In the long time limit, one obtains the constant value for the MSD, given by
\begin{align}
    \lim_{t\rightarrow\infty}\langle x^2(t)\rangle_r&=\lim_{s\rightarrow0}s\,\langle \hat{x}^2(s)\rangle_r = r\, \langle \hat{x}^2(r)\rangle\nonumber\\&=\frac{x_0^2\,r^{\frac{1+\nu}{2}}+\frac{\sigma_x^2}{2\left(\sigma_y^2/2\right)^{\frac{1-\nu}{2}}}}{r^{\frac{1+\nu}{2}}+\frac{\lambda}{\left(\sigma_y^2/2\right)^{\frac{1-\nu}{2}}}}.
\end{align}
For $r=0$ we recover the previous result~(\ref{msd long fractal}) for the case without resetting, and for $\nu=0$ the result~(\ref{msd_reset_in_laplace_append}) for the standard comb with $\mu=0$.

\section{Summary}\label{secSummary}

In this work, we investigated the O-U process in presence of stochastic resetting to the initial position of the particle. We used the Langevin equation approach to perform the numerical simulation and the Fokker-Planck equation to find analytical results for the PDF, NESS and the MSD. We also performed a thorough analysis of the compound effect of the comb geometry and stochastic resetting on the statistical properties of the O-U process. Even though the standard O-U process, with and without resetting, has been considered before, a geometry impact was an open question and the suggested paper on the detailed study of the influence of a comb structure and its anomalous properties can be the answer to the question. We found the corresponding Fokker-Planck equations for the marginal PDFs along the backbone and fingers of the comb in absence and presence of Poissonian resetting. The corresponding mean displacement and the MSD of the particle are calculated exactly by using the one parameter, two parameter and multinomial Mittag-Leffler functions, and the obtained results are confirmed by numerical simulations performed in the framework of the coupled Langevin equations. We also introduced the O-U process on a fractal grid structure, and we have showed that the fractal dimension of the fractal backbone structure has an influence on the PDF and MSD.

In conclusion, we admit that the O–U process is one of several classical approaches used to model interest rates, currency exchange rates, and commodity prices stochastically. The parameter $\mu$ represents the equilibrium or mean value supported by fundamentals; $\sigma$ plays the role of the degree of volatility around it caused by shocks, and $\lambda$ becomes the rate by which these shocks dissipate and the variable reverts towards the mean. However, a plethora of studies \cite{mantegna1995scaling, mantegna1999introduction, bassler2007nonstationary, gopikrishnan1999scaling} show that the distribution of returns $\log[x(t + dt)/x(t)]$ has a sharper maximum and fatter tails, thus further suggesting that a simple O-U trajectory may not be an adequate representation for these types of asset dynamics, due to asymmetries found when comparing its properties with empirical distributions. In addition, an empirical trajectory of interest rates or currency exchange rates may exhibit approximately constant values between two points in time, due to market inactivity. These constant periods can be considered to be trapping of particles, as it is done in physical systems that manifest anomalous diffusion (subdiffusion) \cite{scalas2000fractional,raberto2002waiting}. This empirical investigation represents a potential research avenue for application of the models considered in this work and the further analysis we are leaving for future work.



\begin{acknowledgments}
{The Authors thank Viktor Domazetoski for useful discussions on numerical simulations for the comb structure. PT, PJ, KZ, LK and TS acknowledge financial support by the German Science Foundation (DFG, Grant number ME 1535/12-1). This work is also supported by the Alliance of International Science Organizations (Project No. ANSO-CR-PP-2022-05). AI acknowledges the hospitality at the MPIPKS, Dresden. TS was supported by the Alexander von Humboldt Foundation.}
\end{acknowledgments}

\appendix

\section{Solution of the Fokker-Planck equation for the Ornstein-Uhlenbeck process}\label{appA}

The equation for the standard O-U process is
\begin{align}\label{Fokker_planck_OU_eq} 
\frac{\partial}{\partial t}P_0(x,t)=L_{FP}P_0(x,t),
\end{align}
where
\begin{align}\label{FP_operator_app}
L_{FP}\equiv\lambda\,\frac{\partial}{\partial x}(x-\mu)+\frac{\sigma^2}{2}\,\frac{\partial
^2}{\partial x^2},
\end{align}
is the Fokker-Planck operator. The initial condition is $P(x,t=0)=\delta(x-x_0)$ and zero boundary conditions are chosen at infinity. This equation is solved by the method of characteristics~\cite{Abbott1966} in Fourier space. Equation~(\ref{Fokker_planck_OU_eq}) in Fourier space reads
\begin{align}
\frac{\partial \tilde{P}_0(k,t)}{\partial t}= -\lambda k\, \frac{\partial \tilde{P}_0(k,t)}{\partial k}-\biggr[\frac{k^2 \sigma^2}{2}+i\mu\lambda k\biggr]\tilde{P}_0(k,t).
\end{align}
The  Lagrange–Charpit equations for this equation are
\begin{align}
    \frac{\partial t}{\partial u}=1;\,\,
    \frac{\partial k}{\partial u}=\lambda k;\,\,
    \frac{\partial \tilde{P}_0}{\partial u}=-\biggr[\frac{k^2 \sigma^2}{2}+i\mu\lambda k\biggr]\tilde{P}_0(k,t),
\end{align}
and the parametrization invariant form of the Lagrange–Charpit equations is
\begin{align}\label{Lagrange–Charpit equations}
    \frac{dt}{1}=\frac{dk}{\lambda k}=\frac{d\tilde{P}_0(k,t)}{-\left(\frac{k^2 \sigma^2}{2}+i\mu\lambda k\right)\tilde{P}_0(k,t)}.
\end{align}
From the first two terms with integration we get
\begin{align}
    k=k_0\,e^{\lambda t} \quad \rightarrow \quad  k_0=k\,e^{-\lambda t},
\end{align}
and then again from the last two terms of \ref{Lagrange–Charpit equations} by integrating with separation of the variables we get the expression 
\begin{align}\label{pdf_fourier}
    \tilde{P}_0(k,t)=C\,e^{-\frac{i \mu\lambda k +\frac{\sigma^2k^2}{4}}{\lambda}}=C\,e^{-\frac{i \mu k_0e^{\lambda t} \lambda +\frac{k_0^2 e^{2\lambda t} \sigma^2}{4}}{\lambda}}.
\end{align}
The coefficient $C$ is determined at time $t=0$, when $\hat{P}_0(k,0)=e^{ik_0x_0}$. It follows that $C$ has the form
\begin{align}
    C=e^{ik_0x_0}\,e^{\frac{i\mu \lambda k_0+\frac{\sigma^2 k_0^2}{4}}{\lambda}}.
\end{align}
Inserting the coefficient $C$ in Eq.~(\ref{pdf_fourier}) and exchanging for $k_0=k\,e^{-\lambda t}$ we get the final form of the PDF in Fourier space
\begin{align}
    \hat{P}_0(k,t)=e^{ikx_0\,e^{-\lambda t}}\, e^{-\frac{i \mu k \lambda}{\lambda} (e^{\lambda t}-1)e^{-\lambda t}-\frac{k^2 \sigma^2}{4 \lambda}(e^{2\lambda t}-1)e^{-2\lambda t}}.
\end{align}
By the inverse Fourier transform of the last expression, we get the solution for the PDF of the standard O-U process
\begin{align}\label{pdf_solution_standard_O-U_appen}
    P_0(x,t)=\frac{\exp\left(-\frac{\left[x-x_0e^{-\lambda t}-\mu \left(1-e^{-\lambda t}\right)\right]^2}
    {\frac{\sigma^2}{\lambda} e^{-2\lambda t}\left(e^{2\lambda t}-1\right)}
    \right)}{\sqrt{2\pi\frac{\sigma^2}{2\lambda}e^{-2\lambda t}\left(e^{2\lambda t}-1\right)}
    }.
\end{align}

\section{Hermite function}\label{appHermite}

\begin{figure}[h!]
\includegraphics[width=7.5cm]{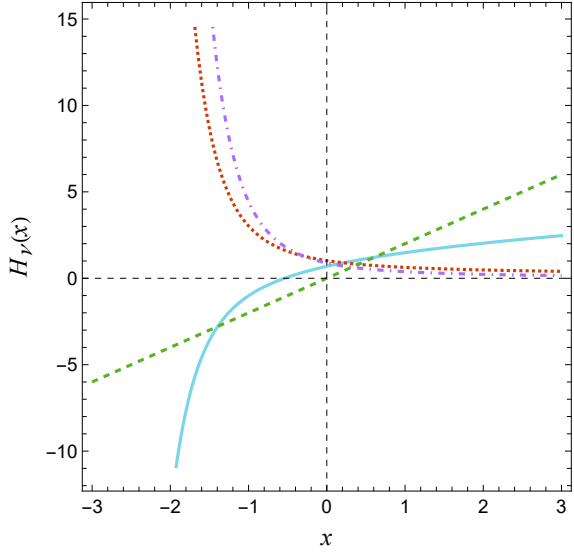}
\caption{Hermite functions ~(\ref{Hermite_functions}) for both positive and negative orders: $\nu=1/2$ (blue solid line), $\nu=-1/2$ (red dotted line), $\nu=1$ (green dashed line) and $\nu=-1$ (purple dot-dashed line).}\label{Hermite functions}
\end{figure}

The solution of the following Hermite differential equation
\begin{align}
w''(z)-2z\,w'(z)+2\nu\, w(z)=0
\end{align}
is given by
\begin{align}
    w(z)=c_1\,H_{\nu}(z)+c_2\,e^{z^2}H_{-\nu-1}(\imath z),
\end{align}
where
\begin{align}\label{Hermite_functions}
    H_{\nu}(z)=2^{\nu}\sqrt{\pi}&\left[\frac{1}{\Gamma\left(\frac{1-\nu}{2}\right)}\,{_1}F_{1}\left(-\frac{\nu}{2},\frac{1}{2},z^2\right)\right.\nonumber\\&\left.-\frac{2z}{\Gamma\left(-\frac{\nu}{2}\right)}\,{_1}F_{1}\left(\frac{1-\nu}{2},\frac{3}{2},z^2\right)\right]
\end{align}
is the Hermite function~\cite{WM_Hermite}\footnote{The Hermite function $H_{\nu}(z)$ is implemented in Wolfram Language as $\mathtt{HermiteH[\nu, z]}.$}, and ${_1}F_{1}(a,b,z)$ is the confluent hypergeometric function. For $\nu=n\in \mathbb{N}$, the Hermite function reduces to the Hermite polynomials, see also Fig.~\ref{Hermite functions}.

The series expansion of the Hermite function for $z\rightarrow0$ is given by~\cite{WM_Hermite}
\begin{align}
    &H_{\nu}(z)=\frac{2^{\nu}\sqrt{\pi}}{\Gamma\left(\frac{1-\nu}{2}\right)}\left[1-\nu\,z^2-\frac{\nu(2-\nu)}{6}z^4+\dots\right]\nonumber\\&-\frac{2^{\nu+1}\sqrt{\pi}}{\Gamma\left(-\frac{\nu}{2}\right)}\,z\left[1+\frac{1-\nu}{3}\,z^2+\frac{(1-\nu)(3-\nu)}{30}z^4+\dots\right].
\end{align}
For $|z|\rightarrow\infty$ one can use the following asymptotic expansion formula~\cite{WM_Hermite}
\begin{align}
    H_{\nu}(z)&\sim (z^2)^{-\frac{\nu}{2}-1} \biggr\{\frac{\sqrt{\pi}\,e^{z^2}\left( \sqrt{z^2} - z\right)}{2\Gamma(-\nu)}\left[1 + O\left(\frac{1}{z^2}\right)\right]\nonumber\\& - 2^{\nu}\sqrt{-z^2}\,(-z^4)^{\nu/2} \Big[\sqrt{-z^2} \,\cos(\nu\pi/2) \nonumber\\&+ z\, \sin(\nu\pi/2)\Big] \left[1 + O\left(\frac{1}{z^2}\right)\right]\biggr\}.
\end{align}

The following formulas hold true for the first derivative of the Hermite function~\cite{WM_Hermite}
\begin{align}
    \frac{\partial}{\partial z}H_{\nu}(z)=2\nu\,H_{\nu-1}(z),
\end{align}
\begin{align}
    \frac{\partial}{\partial z}\left[e^{-z^2}H_{\nu}(z)\right]=-e^{-z^2}H_{\nu+1}(z),
\end{align}
which can be used to obtain the constants in the solutions for the NESS in Eqs.~(\ref{sol pr1}) and~(\ref{sol pr2}).

\section{Fractional calculus and related Mittag-Leffler functions}\label{appB}

The Riemann-Liouville fractional integral of order $\mu>0$ is defined by~\cite{prudnikov2003integrals}
\begin{align}\label{Eq.7}
I_{0+}^{\mu}f(t)=\frac{1}{\Gamma(\mu)}\int_{0}^t\frac{f(t')}{(t-t')^{1-\mu}}\,dt', \quad \Re(\mu)>0,
\end{align}
such that for $\mu=0$ it is 
\begin{align}\label{rl integral mu0}
    I_{0+}^{0}f(t)=f(t).
\end{align}

The Riemann-Liouville fractional derivative of order $0<\mu<1$ is defined as a derivative of the Riemann-Liouville fractional integral of a function~\cite{prudnikov2003integrals},
\begin{align}\label{rl derivative}
    {_{\text{RL}}}D_{t}^{\mu}f(t)&=\frac{d}{dt}I_{0+}^{1-\mu}f(t)\nonumber\\&=\frac{1}{\Gamma(1-\alpha)}\frac{d}{dt}\int_{0}^{t}(t-t')^{-\mu}f(t')\,dt',
\end{align}
while the Caputo fractional derivative of order $0<\mu<1$ is defined as the Riemann-Liouville fractional integral of the first derivative of a function~\cite{prudnikov2003integrals},
\begin{align}\label{caputo derivative}
    {_{\text{C}}}D_{t}^{\mu}f(t)&=I_{0+}^{1-\mu}\frac{d}{dt}f(t)\nonumber\\&=\frac{1}{\Gamma(1-\alpha)}\int_{0}^{t}(t-t')^{-\mu}\frac{d}{dt'}f(t')\,dt'.
\end{align}

The tempered Riemann-Liouville fractional derivative of order $0<\mu<1$ with tempering parameter $r$ is defined by~\cite{mathematics2017,book_ws}
\begin{align}\label{temperedRLderivative}   {_{\text{TRL}}}D_{0+}^{\mu}f(t)=\frac{1}{\Gamma(1-\mu)}\frac{d}{dt}\int_{0}^{t}e^{-r(t-t')}(t-t')^{-\mu}f(t')\,dt'.
\end{align}

The three parameter Mittag-Leffler function (also known as a Prabhakar function) is defined by~\cite{prabhakar1971singular}
\begin{equation}\label{three parameter ML}
E_{\alpha,\beta}^\gamma(z)=\sum_{k=0}^{\infty}\frac{(\gamma)_k}{\Gamma(\alpha k+\beta)}\frac{z^k}{k!},
\end{equation}
where $\beta, \gamma, z \in \mathbb{C}$, $\Re(\alpha)>0$, $(\gamma)_{k}$ is the Pochhammer symbol 
\begin{align}\label{pochhammer symbol ch1}
    (\gamma)_{0}=1, \quad (\gamma)_{k}=\frac{\Gamma(\gamma+k)}{\Gamma(\gamma)}.
\end{align}
It is a generalization of the two parameter Mittag-Leffler function
\begin{align}\label{two parameter ML} E_{\alpha,\beta}^1(z)=\sum_{k=0}^{\infty}\frac{z^{k}}{\Gamma(\alpha k+\beta)}=E_{\alpha,\beta}(z),
\end{align}
and the one parameter Mittag-Leffler function
\begin{align}\label{one parameter ML}    E_{\alpha,1}^1(z)=\sum_{k=0}^{\infty}\frac{z^{k}}{\Gamma(\alpha k+1)}=E_{\alpha}(z).
\end{align}
The associated three parameter Mittag-Leffler function is defined by
\begin{equation}\label{three parameter ML e}
\mathcal{E}_{\alpha, \beta}^{\gamma}(t;\pm\lambda) = t^{\beta-1}\,E_{\alpha, \beta}^{\gamma}(\mp\lambda t^\alpha), 
\end{equation}
with $\min\{\alpha, \beta, \gamma\}>0$, $\lambda \in \mathbb{R}$, and the corresponding Laplace transform
\begin{equation}\label{Laplace ML3_1}
\mathcal{L}\left[\mathcal{E}_{\alpha, \beta}^{\gamma}(t;\pm\lambda)\right]=\frac{s^{\alpha\gamma-\beta}}{(s^\alpha\pm\lambda)^\gamma},
\end{equation}
where $|\lambda/s^{\alpha}|<1$.

The associated multinomial Mittag-Leffler function is defined as follows
\begin{align}\label{associated multinomial ml}
    &\mathcal{E}_{(\alpha_1,\alpha_2,\dots,\alpha_n),\beta}\left(t;\pm\lambda_1,\pm\lambda_2,\dots,\pm\lambda_n\right)\nonumber\\&=t^{\beta-1}\,E_{(\alpha_1,\alpha_2,\dots,\alpha_n),\beta}\left(\mp\lambda_1t^{\alpha_1},\mp\lambda_2t^{\alpha_2},\dots,\mp\lambda_nt^{\alpha_n}\right),
\end{align}
where 
\begin{align}\label{multinomial ml}
&E_{\left(\alpha_{1},\alpha_{2},\dots,\alpha_{n}\right),\beta}\left(z_{1},z_{2},\dots,z_{n}\right)\nonumber\\&=\sum_{k=0}^{\infty}\sum_{l_{1}\geq0,l_{2}\geq 0,\dots,l_{n}\geq0}^{l_{1}+l_{2}+\dots+l_{n}=k}\left(\begin{array}{c l}
    k\\
    l_{1},\dots,l_{n}
  \end{array}\right)\frac{\prod_{i=1}^{n}z_{i}^{l_{i}}}{\Gamma\left(\beta+\sum_{i=1}^{n}\alpha_{i}l_{i}\right)}
\end{align}
is the multinomial Mittag-Leffler function~\cite{lucho_gorenflo}, and $$\left(\begin{array}{c l}
    k\\
    l_{1},\dots,l_{n}
  \end{array}\right)=\frac{k!}{l_1!\,l_2!\,\dots\,l_n!}$$ are the multinomial coefficients. The associated multinomial Mittag-Leffler function can be obtained by the following inverse Laplace transform
\begin{align}
    &\mathcal{L}^{-1}\left[\frac{s^{-\beta}}{1\pm\sum_{j=1}^{n}\lambda_{j}s^{-\alpha_j}}\right]\nonumber\\&=\mathcal{E}_{(\alpha_1,\alpha_2,\dots,\alpha_n),\beta}\left(t;\pm\lambda_1,\pm\lambda_2,\dots,\pm\lambda_n\right).
\end{align}

From the definition of the associated multinomial Mittag-Leffler function (\ref{associated multinomial ml}), one finds that for $n=1$ (i.e., $\lambda_1=\lambda$, $\alpha_1=\alpha$) it corresponds to the associated two parameter Mittag-Leffler function,
\begin{align}
    \mathcal{E}_{(\alpha),\beta}\left(t;\pm\lambda\right)&=\mathcal{L}^{-1}\left[\frac{s^{-\beta}}{1\pm\lambda s^{-\alpha}}\right]\nonumber\\&=\mathcal{L}^{-1}\left[\frac{s^{\alpha-\beta}}{s^{\alpha}\pm\lambda }\right]=t^{\beta-1}\,E_{\alpha,\beta}\left(\mp\lambda_1t^{\alpha_1}\right)\nonumber\\&=\mathcal{E}_{\alpha,\beta}^{1}\left(t;\pm\lambda\right)\equiv \mathcal{E}_{\alpha,\beta}\left(t;\pm\lambda\right).
\end{align}

Moreover, for $n=2$, applying the series expansion approach (see Ref.~\cite{podlubny1998fractional}), we have~\cite{book_ws}
\begin{align}\label{multinomial ml 2parameters}
&\mathcal{E}_{(\alpha_1,\alpha_2),\beta}\left(t;\pm\lambda_1,\pm\lambda_2\right)=\mathcal{L}^{-1}\left[\frac{s^{-\beta}}{1\pm\lambda_1 s^{-\alpha_1}\pm\lambda_2 s^{-\alpha_2}}\right]\nonumber\\&=\mathcal{L}^{-1}\left[\frac{s^{-\beta}}{1\pm\lambda_1 s^{-\alpha_1}}\frac{1}{1\pm\lambda_2 \frac{s^{-\alpha_2}}{1\pm\lambda_1 s^{-\alpha_1}}}\right]\nonumber\\
    &=\sum_{k=0}^{\infty}(\mp\lambda_2)^{k}\,\mathcal{L}^{-1}\left[\frac{s^{-(\alpha_2-\alpha_1)k+\alpha_1-\beta}}{\left(s^{\alpha_1}\pm\lambda_1\right)^{k+1}}\right]\nonumber\\
    &=\sum_{k=0}^{\infty}(\mp\lambda_2)^{k}t^{\alpha_2k+\beta-1}E_{\alpha_1,\alpha_2k+\beta}^{k+1}\left(\mp\lambda_1t^{\alpha_1}\right)\nonumber\\&=\sum_{k=0}^{\infty}(\mp\lambda_2)^{k}\mathcal{E}_{\alpha_1,\alpha_2k+\beta}^{k+1}(t;\pm\lambda_1),
\end{align}
where we also use the Laplace transform (\ref{Laplace ML3_1}) of the associated three parameter Mittag-Leffler function. Thus, the associated multinomial Mittag-Leffler function (\ref{associated multinomial ml}) reduces to infinite series of the associated three parameter Mittag-Leffler functions~(\ref{three parameter ML e}), which is shown to be convergent (see Appendix~C in Ref.~\cite{sandev2011generalized}, and Ref.~\cite{paneva2016bessel}). 

\end{document}